\documentclass[a4paper,11pt]{article}
\usepackage{jcappub} 
\pdfoutput=1
\usepackage{graphicx}
\usepackage{color}
\usepackage{latexsym}
\usepackage{subfigure}
\usepackage{hyperref}
\usepackage{amssymb}
\usepackage{bm}
\usepackage{natbib}
\usepackage{amsmath}
\usepackage{multirow}

\title{21~cm Angular Power Spectrum from Minihalos as a Probe of Primordial Spectral Runnings}

\author[a]{Toyokazu Sekiguchi,}
\author[b]{Tomo Takahashi,}
\author[c]{Hiroyuki Tashiro}
\author[d]{and Shuichiro Yokoyama}

\affiliation[a]{Center for Theoretical Physics of the Universe, Institute for Basic Science, Daejeon 34051, Korea}
\affiliation[b]{Department of Physics, Saga University, Saga 840-8502, Japan}
\affiliation[c]{Department of Physics, Graduate School of Science, Nagoya University, Aichi 464-8602, Japan}
\affiliation[d]{Department of Physics, Rikkyo University, 3-34-1 Nishi-Ikebukuro, Toshima, Tokyo 171-8501, Japan}

\emailAdd{sekiguti@ibs.re.kr, tomot@cc.saga-u.ac.jp, hiroyuki.tashiro@nagoya-u.jp, shuichiro@rikkyo.ac.jp}

\abstract{
Measurements of 21~cm line fluctuations from minihalos have been discussed as a powerful probe
of a wide range of cosmological models. However, previous studies have taken into account only 
the pixel variance, where contributions from different scales are integrated.
In order to sort out information from different scales, we formulate the angular power 
spectrum of 21~cm line fluctuations from minihalos at different redshifts, which can enhance the 
constraining power enormously. By adopting this formalism, we investigate expected constraints on 
parameters characterizing the primordial power spectrum, particularly focusing on the spectral index $n_s$ and its runnings $\alpha_s$ and $\beta_s$.
We show that future observations of 21~cm line fluctuations from minihalos, in combination with cosmic microwave background, 
can potentially probe these runnings as $\alpha_s \sim {\cal O}(10^{-3})$ and $\beta_s \sim {\cal O}(10^{-4})$. 
Its implications to the test of inflationary models are also discussed.
}

\begin{document}
\begin{flushright}
CTPU-17-09
RUP-17-7
\end{flushright}
\maketitle

\section{Introduction}

Current precise measurements of the cosmic microwave background~(CMB) from Planck~\cite{Adam:2015rua,Ade:2015xua} and other 
cosmological observations, such as baryon acoustic oscillation~(BAO)~(see e.g.,~\cite{Alam:2016hwk}), type Ia supernovae (SNe)~(see e.g.,~\cite{Betoule:2014frx}) and so on, 
have unveiled various aspects of the Universe from the very early time to the present.
The very early Universe can be probed by investigating primordial fluctuations whose properties 
can be measured by observing the CMB anisotropies and large scale structures of the Universe. 
Since the primordial fluctuations are considered to be generated during inflation, 
by studying the properties of the primordial fluctuations, we can test
inflationary models and generation mechanisms of these fluctuations. 

The measurement of the CMB by Planck in combination with
other observations has provided strong constraints on inflationary
parameters describing the nature of primordial fluctuations such as their amplitude $A_s$, the spectral index $n_s$,
the tensor-to-scalar ratio $r$ and the non-linearity parameter $f_{\rm NL}$~\cite{Ade:2015lrj}.
However, we are still far from specifying the model of inflation 
and more observational information is required to fully understand the
very early epoch of the Universe.

Among the inflationary parameters, the tensor-to-scalar ratio is one of
the promising quantity which, in the near future, can be
probed more accurately from on-going and planned CMB B-mode polarization 
experiments~(see, e.g.,~\cite{Huang:2015gca,Errard:2015cxa,Abazajian:2016yjj,Alonso:2016xft,Barron:2017kuo}). Since the tensor-to-scalar ratio is directly related to the energy scale of inflation, 
its information would give important implications to the models of inflation. 
Another property of primordial fluctuations worth investigating is their scale-dependence, 
which is usually represented by the spectral index $n_s$. Although $n_s$ at a reference wavenumber on large scales is accurately determined by 
Planck data \cite{Ade:2015lrj}, it also depends on the scale in general. The scale-dependence of $n_s$ is commonly denoted as the running $\alpha_s$,
which is defined as $\alpha_s = d n_s / d \ln k$. Furthermore, $\alpha_s$ also generally depends on the scale, which is referred as the running of the running 
$\beta_s = d^2 n_s / d \ln^2 k$. Compared to $n_s$,  the current constraints
on these runnings are not so severe even with the Planck data~\cite{Ade:2015lrj}.
However, these can be probed more precisely by future observations on smaller scales such
as 21 cm line fluctuations
from the intergalactic medium (IGM)~\cite{Kohri:2013mxa,Munoz:2016owz},  the CMB spectral $\mu$ distortion~\cite{Dent:2012ne,Cabass:2016giw} and galaxy surveys~\cite{Munoz:2016owz}.

We in this paper argue that future observations for the angular power
spectrum of 21~cm line fluctuations from minihalos,
which are virialized objects with the virial temperature
$T<10^4$~K,  would be a useful probe of the shape of the primordial power spectrum.
Since the virial temperature of minihalos is not high enough to cause
the effective collisional ionization, the inside of a minihalo is filled
with dense neutral gas. Therefore, the existence of minihalos can induce
the additional 21~cm line signals~\cite{Iliev:2002gj,Iliev:2002ms}.
Since the abundance of minihalos depends on the matter
density fluctuations at $20~{\rm Mpc}^{-1} < k < 500~{\rm Mpc}^{-1}$, 
the 21~cm signatures from minihalos have been discussed as a probe of various cosmological models, particularly on small scales.
In Ref.~\cite{Shimabukuro:2014ava}, the potential to probe small-scale
primordial fluctuations has been investigated
by using absorption features produced by minihalos in the continuum spectrum of
the radio sources at high redshifts.

The clustering of minihalos has also been studied
in the context of warm dark matter~\cite{Sekiguchi:2014wfa}, 
isocurvature fluctuations~\cite{Sekiguchi:2013lma,Takeuchi:2013hza},
primordial non-Gaussianity~\cite{Chongchitnan:2012we}, cosmic
strings~\cite{Tashiro:2013xra}
and so on, because the clustering
can enhance the fluctuations of 21~cm line signals related
to the matter density fluctuations.
In other words, minihalos are biased tracers of the matter density fluctuations.
Once we measure how much the 21~cm line fluctuations are biased by minihalos,
we can probe 
matter fluctuations on much smaller scales
compared with CMB or large scale structure observations.
However, in these previous studies, only the pixel variance of 21~cm line fluctuations has been
focused on to investigate the feasibility of future 21~cm line observations.
Although the pixel variance would be  easy to obtain from  actual observational 
data, 
it cannot tell us the contributions from different scales. 
Hence, instead of the pixel variance, we discuss the 21~cm line signals from
minihalos in terms of the angular power spectrum 
which can probe the contributions from different scales.
We show that the next generation 21~cm survey 
such as Square Kilometre Array~(SKA)~\cite{ska}
and Fast Fourier Transform Telescope~(FFTT)~\cite{Tegmark:2008au} can
measure the scale-dependence very accurately,
which will be very helpful to 
differentiate models of inflation.

The structure of this paper is as follows. In the next section, we describe the formalism 
to calculate the angular power spectrum of 21~cm line fluctuations from minihalos. 
Then in Section~\ref{sec:fisher}, we discuss the Fisher matrix for the angular power spectrum of 21~cm line fluctuations
as well as that for CMB. In Section~\ref{sec:application},   we estimate expected 
constraints on the power spectrum of primordial perturbations, focusing on the runnings of the spectral index,
 assuming future 21~cm surveys including SKA and FFTT
in combination with CMB observations. Implications for specific models for inflation are also discussed here.
We conclude in the final section. In Appendix \ref{app:higher_power_slow_roll}, 
we present general expressions for the spectral parameters for 
single-field inflation models and multi-field ones with the inflaton and spectator fields.
In Appendix \ref{app:higher_running}, constraints on higher order spectral runnings are given.

\section{Angular power spectrum of 21~cm line fluctuations from minihalos}
\label{sec:formalism}

Minihalos are neutral virial objects which are enough dense to decouple
the spin temperature of neutral hydrogen from the CMB temperature and can produce the observable 21~cm signal.
Since minihalos are biased objects of the matter density fluctuations,
the abundance of minihalos also fluctuates.
As a result, the fluctuations of 21~cm signals due to minihalos arise.
Here, following the treatment of the previous works~\cite{Iliev:2002gj,Iliev:2002ms,Chongchitnan:2012we,
Tashiro:2013xra,Takeuchi:2013hza,Sekiguchi:2013lma,Sekiguchi:2014wfa},
we derive the angular power spectrum of 21 cm line fluctuations from
minihalos.

We are interested in 21~cm line fluctuations on scales which are much larger than the typical formation scale of minihalos.
In other words, the angular resolution of an observation is much larger than the size of individual minihalos.
In this case, there exist multiple minihalos
in a beam size of the observation.
The observable of 21~cm measurements is written in terms of the
differential brightness temperature, which represents the deviation of
the brightness temperature from
the CMB temperature.
The mean differential brightness temperature due to minihalos at
redshift $z$ can be obtained from~\cite{Iliev:2002gj}
\begin{equation}
\overline{\Delta T_b}(z)
=
c\frac{(1+z)^4}{H(z) \nu_0}\int^{M_{\rm max}}_{M_{\rm min}}
dM~\frac{dn}{dM} \Delta\nu_{\rm eff}  A\langle \delta T_b \rangle,
\label{eq:dtb}
\end{equation}
where $\nu_0$ is the frequency corresponding to 21~cm, $\Delta \nu_{\rm
eff} =[ \phi(\nu_0)(1+z)]^{-1}$ with the intrinsic line profile
$\phi(\nu_0)$ of a minihalo, $dn/dM$ is a mass function of minihalos, 
$A$ is the cross-section of a minihalo, and $\langle \delta T_b \rangle$ is the
typical brightness temperature of a minihalo averaged over the
cross-section $A$. We consider that minihalos are in the mass range
between $M_{\rm max}$ and $M_{\rm min}$. We set $M_{\rm max}$ and
$M_{\rm min}$ to the virial mass with the virial temperature $T_{\rm
vir} =10^4$\,K and the Jeans mass, respectively. As $dn/dM$, we adopt the Press-Schechter
mass function.
The nature of minihalos, e.g., the profiles of the gas density and
pressure, determines $A$, $\phi(\nu_0)$ and $\langle \delta T_b
\rangle$. Here we use a truncated isothermal sphere
as the model of a minihalo, which depends on the minihalo mass $M$~\cite{Shapiro:1998zp}. 
Minihalos are clustered, depending on the underlying density fluctuations.
Therefore, 
21~cm line fluctuations from minihalos at a redshift $z$ in the line-of-sight direction $\hat n$
should be written as (neglecting the redshift space
distortions)~\cite{Iliev:2002gj,Sekiguchi:2013lma,Sekiguchi:2014wfa}
\begin{equation}
\delta  \Delta T_b(\hat n, z)  =
\overline{\Delta T_b}(z)
\beta(z)
\delta(\vec x=r(z)\hat n, z),
\label{eq:deltaDelta}
\end{equation}
where $\beta(z)$ is the effective bias of minihalos,
$\delta(\vec x, z)$ represents the matter density fluctuations at the comoving coordinates $\vec x$ and redshift $z$,
and $r(z)$ is the comoving distance from us to the redshift $z$.
The effective bias $\beta(z)$ is given
as
\begin{equation}
\beta(z)\equiv \frac{\int^{M_{\rm max}}_{M_{\rm min}}  dM \frac{dn}{dM} \mathcal F(z,M) b(M,z)}
{\int^{M_{\rm max}}_{M_{\rm min}}  dM \frac{dn}{dM} \mathcal F(z,M)},
\label{eq:ave-bias}
\end{equation}
where $\mathcal F= \langle \delta T_b \rangle A \sigma_V$ with the velocity
dispersion of a minihalo $\sigma_V$ and
$b(M,z)$ being the bias of minihalos with
mass $M$ for which 
we adopt one obtained in Ref.~\cite{Mo:1995cs}.  
We refer the readers to Refs.~\cite{Iliev:2002gj,Sekiguchi:2013lma,Sekiguchi:2014wfa} 
for the detailed derivation of $\overline{\Delta T_b}(z)$ and $\beta(z)$.

Minihalos discussed in this paper are collapsed objects and the corresponding scales of matter density
fluctuations are at $20~{\rm Mpc}^{-1} < k < 500~{\rm Mpc}^{-1}$~\cite{Sekiguchi:2014wfa}.
Therefore, the bias~(clustering) as well as the mean number density of minihalos depends on the fluctuations at these scales.
As shown in Eq.~(\ref{eq:deltaDelta}) with Eq.~(\ref{eq:ave-bias}), the
clustering enhances the signals of 21~cm fluctuations over all
scales. As a result, the measurement of the 21 cm line fluctuation amplitude
provides useful information about the matter density fluctuations on small
scales where minihalos form.

\begin{figure}
  \begin{center}
    \hspace{0mm}\scalebox{.8}{\includegraphics{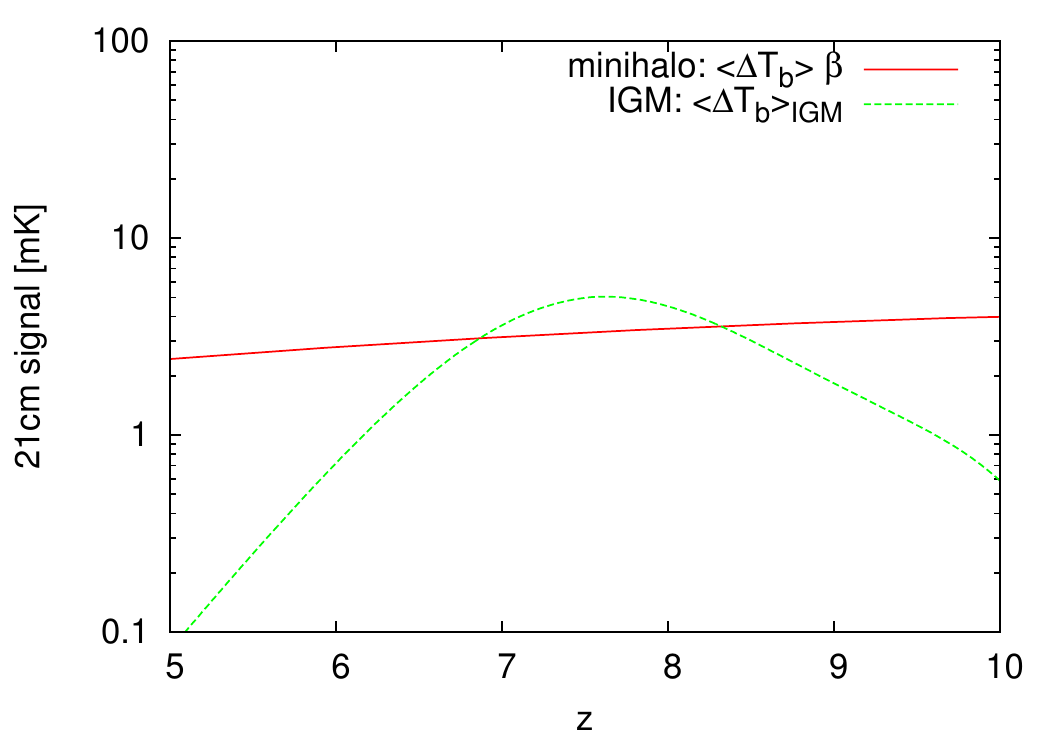}}
  \end{center}
  \vspace{10mm}
  \caption{Comparison of the signals in  21 cm line fluctuations from minihalos
 (red solid line) and the IGM (green dashed line) as functions of redshift $z$. 
 For the IGM signal, we have assumed a rapid reionization with the optical depth  $\tau_{\rm reion}=0.058$ and  the width of the  duration $\Delta z = 1$ in accordance with the recent result \cite{Adam:2016hgk}.
  }
  \label{fig:dTb}
\end{figure}

Although  we focus on the 21 cm signals from minihalos in this paper,
the IGM also contributes to the signal, 
which can be written as
\begin{equation}
\label{eq:Tb_IGM}
\Delta {T_b}_{\rm IGM} \approx 28{\rm mK} \sqrt{\frac{z+1}{10}} x_{\rm HI}(z)(1+\delta)(1-T_{\rm CMB}/T_s),
\end{equation}
where $x_{\rm HI}(z)$ and $T_s$ are the neutral fraction 
and the spin temperature of the IGM, respectively~\cite{Madau:1996cs}.
Which contribution dominates the signal, from minihalos or the IGM, depends on
the thermal and ionization history of the IGM.
In general, the reionization process proceeds faster in the IGM than in
minihalos~\cite{Shapiro:2003gxa}.
This is because minihalos are 
$\mathcal O(100)$ times denser than the IGM. 
Accordingly much more ionizing background photons are required to ionize
minihalos in comparison with the IGM. 
In this paper, for simplicity, we assume that minihalos are not ionized at all until the completion of reionization.
Under this assumption, Fig.~\ref{fig:dTb} compares the signal of the 21 cm
line fluctuations from minihalos, $\overline{\Delta T_ b}(z) \beta$, 
with that of the IGM $\overline{\Delta T_b}_{\rm IGM}$ given in Eq.~\eqref{eq:Tb_IGM}. 
In the figure,  we assume a rapid reionization with the optical depth  $\tau_{\rm reion}=0.058$ and  the width of the  duration $\Delta z = 1$ in accordance with the recent Planck result~\cite{Adam:2016hgk}
for $x_{\rm HI}(z)$ to calculate the signal from the IGM.
We obtain the spin temperature, taking the simple model of the IGM gas
temperature evolution in which the gas temperature is
proportional to $x_{\rm HI}$ during the reionization and reaches
$10^4~$K when the reionization completes~\cite{Barkana:2000fd}.
The figure shows that the signal from minihalos can surpass that from IGM in almost overall redshifts.
Although, as mentioned above, the signal from the IGM depends on the reionization history, 
we have also checked other several models and found that, even if we change the reionization history, 
the signal from minihalos dominates for some redshift range.
Therefore, in the following analysis, we neglect the contribution from the IGM.
We also note that the contribution from minihalos is dominant in the end of the dark
ages~($15< z < 40$). At this epoch, the temperature of the IGM is so
low that the spin temperature is almost same as the CMB
temperature. Therefore, the signals from the IGM are
suppressed. However, as we will discuss later, the observational noise
becomes large as the observation redshift increases.
Hence  we consider the signals only from $z<20$ in our analysis.

Given a redshift $z$, we consider maps of the 21~cm line fluctuations $\delta  \Delta T_b(\hat n,z) $
and matter ones $\delta(r(z)\hat n, z)$ in the sky. They can be
transformed into the spherical harmonic coefficients as
\begin{eqnarray}
a^{\rm (21cm)}_{lm}(z)= \int d\hat n\,
\delta \Delta T_b(\hat n, z) Y^*_{lm}(\hat n), \\
a^{\rm (matter)}_{lm}(z)= \int d\hat n\, 
\delta(r(z)\hat n, z) Y^*_{lm}(\hat n).
\end{eqnarray}
Due to the statistical isotropy of the matter fluctuations, 
the angular power spectrum of the 21~cm line 
fluctuations from minihalos should be given as 
\begin{equation}
C^{\rm (21cm)}_l(z,z')=
\overline{\Delta T_b}(z)\overline{\Delta T_b}(z')
\beta (z)\beta (z')
C^{\rm (matter)}_l(z,z'), 
\end{equation}
where $C^{\rm (21cm)}_l(z,z')$ and 
$C^{\rm (matter)}_l(z,z')$ are respectively the angular power spectra of 21cm line fluctuations and the matter ones
between redshifts $z$ and $z'$, whose definitions are 
\begin{eqnarray}
\langle a^{\rm (21cm)}_{lm}(z) a^{\rm (21cm)*}_{l'm'}(z')\rangle&=&
C^{\rm (21cm)}_l(z,z') \delta_{ll'}\delta_{mm'}, \\
\langle a^{\rm (matter)}_{lm}(z) a^{\rm (matter)*}_{l'm'}(z')\rangle&=&
C^{\rm (matter)}_l(z,z') \delta_{ll'}\delta_{mm'}. 
\end{eqnarray}
The angular power spectrum of matter fluctuations $C_l^{\rm (matter)}$ 
can be related to the matter power spectrum $P(k)$ as 
\begin{equation}
C^{\rm (matter)}_l(z,z')
=D(z)D(z')\int \frac{k^2dk}{2\pi^2} P(k) j_l(kr(z)) j_l(kr(z')), 
\label{eq:nodist}
\end{equation}
where $D(z)$ is the growth factor relative to a reference redshift $z_{\rm ref}$  (i.e., $\delta(k,z) = D(z) \delta (k,z_{\rm ref})$ with $\delta (k, z)$ being the matter 
fluctuations in the $k$-space) and
the linear matter power spectrum at a reference redshift $z_{\rm ref}$ is denoted as $P(k)$.
$j_l$ is the spherical Bessel function. 
In the redshift range we consider in the following analysis, the Universe is well approximated as matter-dominated, 
in which $D(z)$ is given by $D(z) \propto a(z)$, with $a(z)$ being the scale factor at $z$.

So far we have neglected the redshift space distortions due to the peculiar velocity of minihalos.
Although this effect has not been taken into account in previous studies~\cite{Iliev:2002gj,Iliev:2002ms,Chongchitnan:2012we,
Tashiro:2013xra,Takeuchi:2013hza,Sekiguchi:2013lma,Sekiguchi:2014wfa},
it can be dominant especially in correlations between different
redshifts.
The redshift space distortions are classified largely into two types.
One is the linear effect known as the Kaiser effect
while the other is the nonlinear effect called as the Fingers-of-God (FoG) effect~\cite{Kaiser:1987qv}.
Since we are focusing on  the Universe at high redshifts, we expect that the redshift space distortions are weak and 
can be well-approximated within the framework of the linear perturbation
theory. Thus we neglect the FoG effect. 
Including the Kaiser
effect~\cite{Kaiser:1987qv}, we can rewrite Eq.~(\ref{eq:deltaDelta}) as
\begin{equation}
\delta \Delta T_b(\hat n, z)=
\overline{\Delta T_b}(z)
\left[\beta(z)+f(z)\mu^2 \right]
\delta(\vec x=r(z)\hat n, z), 
\end{equation}
where $\mu=\hat k\cdot \hat n$ is the cosine between the wave vector of perturbations $\vec k$ 
and line-of-sight direction $\hat n$, 
and $f=d\ln D/d\ln a$ is the growth rate.
Note that in the matter domination epoch, $f=1$.
After taking account of the redshift space distortions, 
Eq.~\eqref{eq:nodist} should then be replaced with
\begin{eqnarray}
C^{\rm (matter)}_l(z,z') 
&=&D(z)D(z')\int \frac{k^2dk}{2\pi^2} P(k) \\
&&\times \left[j_l(kr(z))-\frac{f(z)}{\beta(z)}j^{''}_l(kr(z))\right]
\left[j_l(kr(z'))-\frac{f(z')}{\beta(z')}j^{''}_l(kr(z'))\right], \notag
\label{eq:withdist}
\end{eqnarray}
where $j_l^{''}$ denotes a second derivative with respect to its argument and it should not be confused with 
a prime attached to $z$.

\begin{figure}
  \begin{center}
    \begin{tabular}{cc}
    \includegraphics{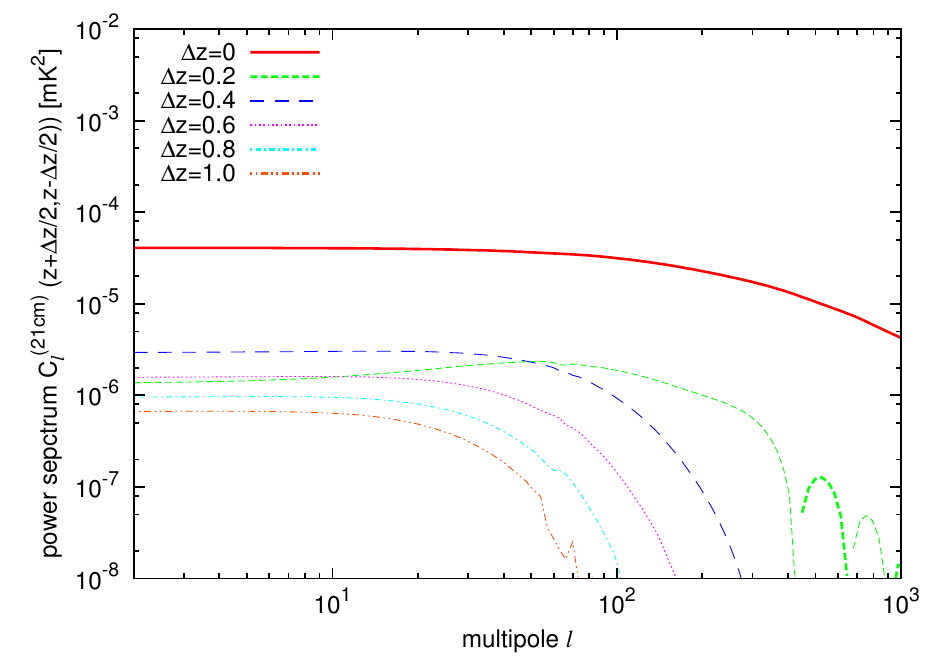}
    \end{tabular}
  \end{center}
  \vspace{10mm}
  \caption{Shown are the angular power spectra of minihalo 21cm line fluctuations $C^{\rm (21cm)}_l(z+\frac
  {\Delta z}2,z-\frac{\Delta z}2)$ with the central redshift $z=5$.   Thick and thin lines show positive and negative values,
 respectively. 
 The cosmological parameters are taken to be the mean values of the analysis for a power-law $\Lambda$CDM model from Planck 2015 TT,TE,EE+lowP+lensing+ext \cite{Ade:2015xua}:
 CDM density $\omega_c (=\Omega_bh^2) =0.1188$,  baryon density $\omega_b(=\Omega_ch^2) =0.0223$, 
the reduced Hubble parameter $h=0.6774$, the reionization optical depth $\tau_{\rm reion}=0.066$, the amplitude and the spectral index of primordial power spectrum (at $k_0=0.05~{\rm Mpc}^{-1}$) $A_s=2.141\times10^{-9}$
and $n_s=0.9667$, respectively.
  }
  \label{fig:cl21cm}
\end{figure}

In Fig.~\ref{fig:cl21cm}, we plot the angular power spectra $C^{\rm (21cm)}_l (z, z')$
around a central redshift $z=5$ as an example.
The cosmological parameters we assume to calculate $C^{\rm (21cm)}_l (z, z')$ are 
given in the caption of Fig.~\ref{fig:cl21cm}.
From the figure, one can see that
the amplitude of the power spectra becomes maximum when the
redshift difference, $\Delta
z\equiv |z-z'|$, vanishes. Then the amplitude 
drops as $\Delta z$ increases, with $l$ being fixed.
The correlation at the same redshift, $C^{\rm (21cm)}_l(z,z)$,
is almost constant at large angular scales $\ell \lesssim 100$.
In other words,
the correlations between different 
redshifts and different pixels are not very large.
This result supports the previous analysis of 21~cm signals from minihalos in which
only pixel variance at the same redshift bins
are taken into account while the covariance is neglected.
However, at smaller angular scales $l\gtrsim 100$,
the angular spectrum deviates from the white spectrum $C_l=$const., 
which indicates nonzero covariance between different pixels.
In addition, with small but nonzero redshift differences $\Delta z\sim 0.1$,
the correlations between different redshifts give non-negligible
amplitude. 
This suggests that, in order to optimally exploit cosmological information
in 21~cm line fluctuations from minihalos, we need to take into account
the correlation of their fluctuations at different redshifts, too.

\section{Fisher matrix analysis}
\label{sec:fisher}

Now we discuss the Fisher matrix for $C^{\rm (21cm)}_l (z,z')$.
To calculate the Fisher matrix,
we adopt the following observational noise for 21cm line fluctuations.
For simplicity, we take the assumption that the noise is white both in angular and 
frequency domains.
Assuming the isotropic Gaussian beam transfer function,
we can write the noise power spectrum as~\cite{Knox:1995dq}
\begin{equation}
N^{\rm (21cm)}_{l,ij}=\delta_{ij}{\varsigma_i}^2\Delta\theta_i^2
\exp\left[l(l+1)\frac{\theta_i^2}{8\ln2}\right],
\label{eq:ncl} 
\end{equation}
where $\varsigma_i$ is the noise root-mean-squared per pixel
and $\theta_i$ is the beam width. 
Following Ref.~\cite{Furlanetto:2006jb}, 
we approximate $\varsigma_i$ as
\begin{equation}
\varsigma_i=20\mbox{mK}\left(\frac{A_{\rm tot}}{10^4 \mbox{m}^2}\right)^{-1}
\left(\frac{\Delta \theta_i}{10^\prime}\right)^{-2}
\left(\frac{1+z_i}{10}\right)^{4.6}
\left(\frac{\Delta \nu_i}{\mbox{MHz}}\frac{t}{100\mbox{h}}\right)^{1/2},
\label{eq:noise}
\end{equation}
from which one can see that the noise increases as the redshift becomes higher.

Then according to Ref.~\cite{Tegmark:1996bz}, the Fisher matrix based on a measurement of 
power spectrum should be given by
\begin{equation}
F^{\rm (power)}_{ab}
=\sum_l\frac{2l+1}2
{\rm Tr}\left[
{\mathbf C^{\rm (21cm)}_l}^{-1} \frac{\partial\mathbf C^{\rm (21cm)}_l}{\partial p_a}
{\mathbf C^{\rm (21cm)}_l}^{-1} \frac{\partial\mathbf C^{\rm (21cm)}_l}{\partial p_b}
\right],
\label{eq:Fisher_power}
\end{equation}
where $p_a$ ($p_b$) indicates a cosmological parameter and $[\mathbf C_l]^{\rm (21cm)}_{ij} \equiv C^{\rm (21cm)}_{l,ij}+N^{\rm (21cm)}_{l,ij} $ is the covariant matrix for 
21 cm observations.

We again note that we in this paper neglect the contribution from IGM 
in order to focus on the potential of the measurement of 21cm signals
from minihalos.
This treatment is also motivated from the fact that, although the IGM component is also expected to contribute to the covariance matrix in Eq.~\eqref{eq:Fisher_power}, 
its contribution would be subdominant in broad redshift range as shown in Fig.~\ref{fig:dTb}.

In a similar fashion, we can define the Fisher matrix from CMB power spectrum measurements:
\begin{equation}
F^{\rm (CMB)}_{ab}
=\sum_l\frac{2l+1}2
{\rm Tr}\left[
{\mathbf C^{\rm (CMB)}_l}^{-1} \frac{\partial\mathbf C^{\rm (CMB)}_{I}}{\partial p_a}
{\mathbf C^{\rm (CMB)}_l}^{-1} \frac{\partial\mathbf C^{\rm (CMB)}_{l}}{\partial p_b}
\right],
\label{eq:Fisher_CMB}
\end{equation}
where $[\mathbf C_{l}]^{\rm (CMB)}_{PQ}\equiv C^{\rm (CMB)}_{l,PQ}+N^{\rm (CMB)}_{l,PQ}$ is the 
covariance matrix of measured CMB anisotropies, with the subscript $P (Q)$ indicating the temperature or E-mode polarization.
Following Ref.~\cite{Knox:1995dq}, we approximate the CMB noise power
spectrum $N^{\rm (CMB)}_{l,PQ}$  as
\begin{equation}
N_{l,PQ}=\delta_{PQ}
\theta_\mathrm{FWHM}^2\sigma_P^2
\exp\left[l(l+1)
\frac{\theta_\mathrm{FWHM}^2}{8\ln2}
\right],
\end{equation}
where $\theta_\mathrm{FWHM}$ are 
the full width at half maximum of the Gaussian
beam, and $\sigma_P$ is the root-mean-square 
of the instrumental noise par pixel.

\section{Constraints on spectral runnings of primordial power spectrum}
\label{sec:application}

In order to demonstrate the potential of the angular power spectrum of 21~cm line fluctuations from minihalos,
here we present expected constraints on the power spectrum of primordial curvature perturbation, particularly focusing on 
the spectral index $n_s$ and its runnings $\alpha_s, \beta_s$.

\subsection{Runnings of the spectral index}

Conventionally the scale-dependence of the primordial power spectrum is given by the power law form with the spectral index $n_s$, which is often 
assumed to be constant in scale (wavenumber). 
However, the spectral index can generally depend on the scale and such scale-dependence could give us detailed information 
on the primordial power spectrum ${\cal P}_s (k) := k^3 P_\zeta (k) / 2 \pi^2$, where $P_\zeta (k)$ is defined as
\begin{equation}
\langle \zeta ({\bm k}) \zeta ({\bm k}') \rangle = (2 \pi)^3 \delta^{(3)} ({\bm k} + {\bm k}') P_\zeta(k),
\end{equation}
with $\zeta$ being the primordial curvature perturbations. Here, $\delta^{(3)} ({\bm k} + {\bm k}')$ is a 3-dimensional Dirac's delta function
and $P_\zeta (k)$ determines the initial condition of the linear matter power spectrum denoted by $P(k)$ through the Poisson equation.
By taking into account the scale-dependence of the spectral index, we can perturbatively write ${\cal P}_s$ as
\begin{equation}
\mathcal P_s(k)=A_s\left(\frac{k}{k_0}\right)^{n_s-1+\frac12 \alpha_s \ln\left( k/k_0 \right)+\frac{1}{3!} \beta_s \ln^2 \left(k/ k_0 \right)},
\label{eq:pow}
\end{equation}
where $A_s$, $k_0$ and  $n_s$ are respectively the amplitude, the pivot scale and the 
 spectral index,
and we have expanded ${\cal P}_s$ in terms of $\ln k$ up to the 2nd order.
The expansion coefficients in the 1st and 2nd orders are denoted as $\alpha_s (\equiv d n_s / d\ln k)$ and $\beta_s  (\equiv d^2 n_s / d\ln^2 k)$, which we call  
the running and the quadratic running of $n_s$,  respectively. 
In the framework of the slow-roll inflation, the runnings such as $\alpha_s$ and $\beta_s$ can be explicitly written down by using
the so-called slow-roll parameters. We provide those expressions in Appendix \ref{app:higher_power_slow_roll}.
In principle, we can expand ${\cal P}_s$ up to arbitrarily higher orders and hence we also, in Appendix \ref{app:higher_power_slow_roll},
give expressions for higher order runnings in terms of the slow-roll parameters not only for the single-field case but also for the multi-field case.

\subsection{Forecasts based on the Fisher matrix analysis}

Let us investigate the expected constraints on primordial power spectrum  in future 21cm line observations, 
especially focusing on the parameters $n_s$, $\alpha_s$ and $\beta_s$. The  determination of  these parameters requires
precise measurements of cosmological perturbations over a wide range of scales, which can be achieved by 
combining observations of the CMB and 21~cm line fluctuations.
For 21~cm line observations, we in this paper adopt the specifications of SKA~\cite{ska} and 
FFTT~\cite{Tegmark:2008au}.
In addition to 21~cm line fluctuations from minihalos, we also combine 
CMB observations with the expected sensitivities of Planck \cite{Planck:2006aa} and COrE~\cite{core}\footnote{
While the survey parameters we adopt here are somewhat different from those in the most recent proposal of the COrE mission,
our results do not differ significantly.
}.
The survey parameters we adopt are summarized in Tables~\ref{tab:spec_21cm} and~\ref{tab:spec_cmb}.
In our analysis, we assume a flat $\Lambda$CDM model and the pivot scale $k_0$ is fixed to $0.05$~Mpc$^{-1}$ as in the Planck analysis.
In addition to $n_s, \alpha_s$ and $\beta_s$, we also include the following parameters in the Fisher matrix: the reduced Hubble parameter $h$,
baryon and CDM densities $\omega_b$ and $\omega_c$,  the reionization optical depth $\tau_{\rm reion}$, and the amplitude of the 
primordial power spectrum  $A_s$.
The fiducial parameters are assumed to be the same as the ones used in Fig.~\ref{fig:cl21cm}.

\begin{figure}
  \begin{center}
    \begin{tabular}{cc}
    \hspace{-5mm}\scalebox{.8}{\includegraphics{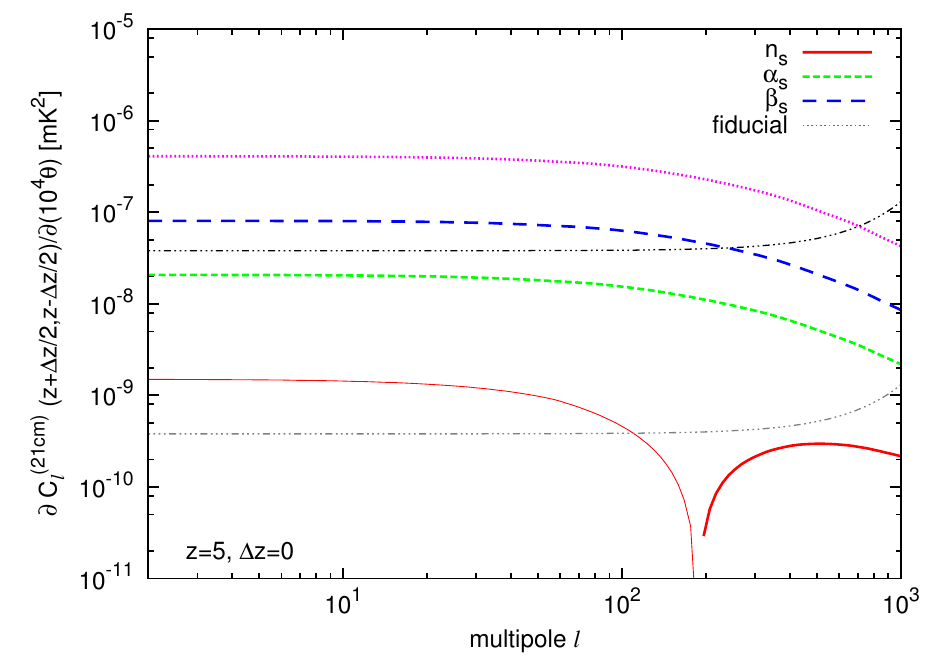}} &
    \hspace{-8mm}\scalebox{.8}{\includegraphics{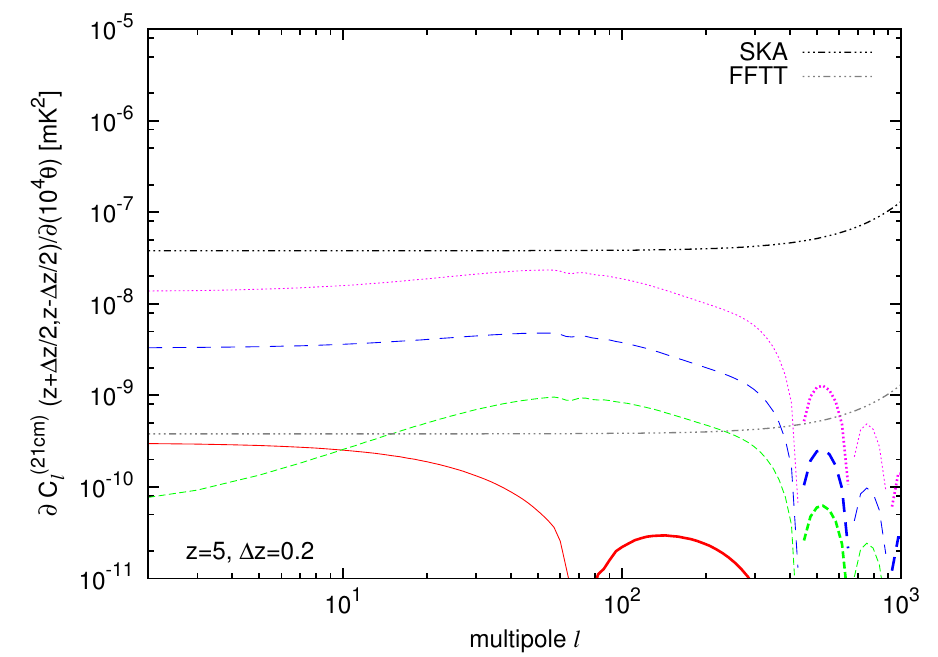}} \\
    \hspace{-5mm}\scalebox{.8}{\includegraphics{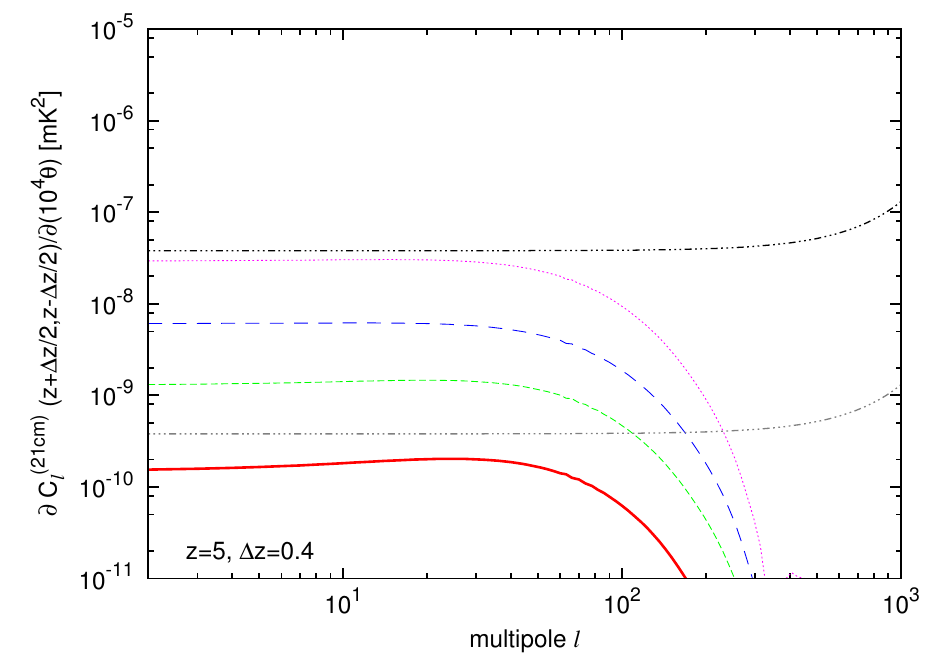}} &
    \hspace{-8mm}\scalebox{.8}{\includegraphics{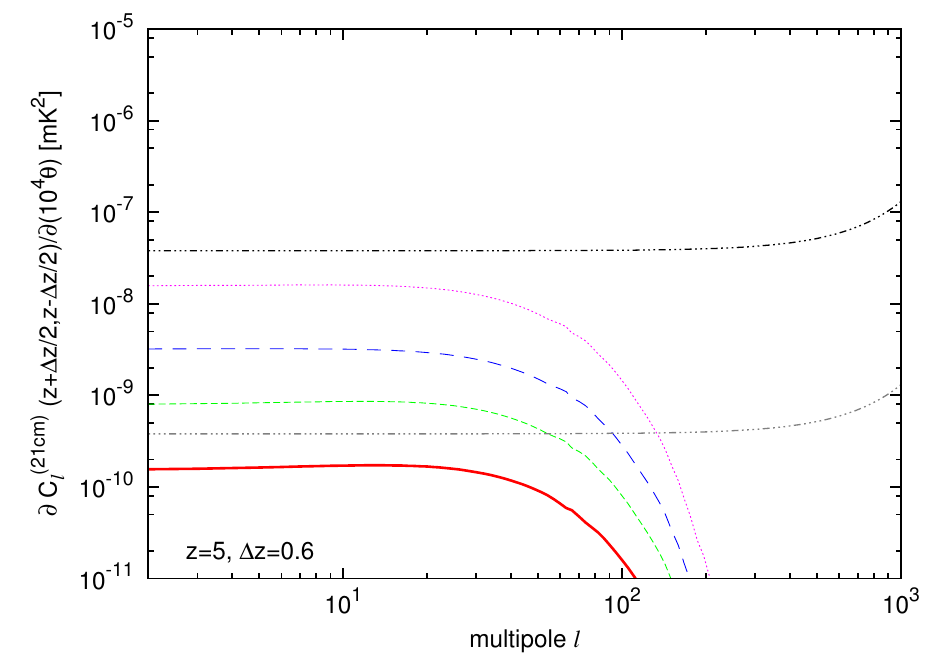}} 
    \end{tabular}
  \end{center}
  \vspace{10mm}
  \caption{Dependence of $C_l^{\rm (21cm)}$ on the parameters of the primordial power spectrum.
  Plotted are the derivative of $C_l^{\rm (21cm)}$ with respect to $\theta = n_s$ (red), $\alpha_s$ (green) and $\beta_s$ (blue),  
divided by $10^4$ for visualization purpose. We also show the fiducial $C_l^{\rm (21cm)}$ (divided by $10^2$) for comparison (purple).
The noise power spectra for SKA and FFTT are also plotted (black).
  The central redshift is fixed to $z=5$ and the redshift difference is varied as 0 (top-left), 0.2 (top-right), 0.4 (bottom-left), 0.6 (bottom-right).
    }
  \label{fig:dcl21cm}
\end{figure}

Fig.~\ref{fig:dcl21cm} shows 
derivatives of $C_l^{\rm (21cm)}$ with respect to parameters 
$n_s$ (red), $\alpha_s$ (green), $\beta_s$ (blue), which captures the dependence of 
$C_l^{\rm (21cm)}(z,z')$ on these parameters.
For reference, the fiducial $C_l^{\rm (21cm)}$ is also depicted (purple).
When $C_l^{\rm (21cm)}$ is highly dependent on a parameter, the derivative
of $C_l^{\rm (21cm)}$
becomes large in the figure.
Here, fixing the central redshift $(z+z')/2=5$, we take $z-z'=$0~(top-left), 0.2~(top-right),
0.4~(bottom-left), 0.6~(bottom-right).
From the figure, we can see the following tendencies:
For $\Delta z = 0$ (top-left), $n_s$ affects  $C_l^{\rm (21cm)}$ in a scale-dependent way, however 
$\alpha_s$ and $\beta_s$ only changes the overall amplitude of $C_l^{\rm (21cm)}$. 
For $\Delta z = 0.2$ (top-right), $n_s$ and $\alpha_s$ can both affect  $C_l^{\rm (21cm)}$ scale-dependently, but 
$\beta_s$ only gives the change in the overall amplitude.
For $\Delta z = 0.4$ (bottom-left) and $0.6$ (bottom-right), all parameters only affect the amplitude of  $C_l^{\rm (21cm)}$ up to $l \sim {\cal O}(100)$.
These trends can be understood as follows.
The spectral index $n_s$ (and the running $\alpha_s$ in some cases) can affect the matter fluctuations $P(k)$ directly on (observable) large scales,
which can affect $C_l^{\rm (21cm)}$ in a scale-dependent manner. 
On the other hand, the quadratic running parameter
$\beta_s$ affects $C_l^{\rm (21cm)}$ largely only through the overall amplitude
since this parameter changes $C_l^{\rm (21cm)}$ mainly through the minihalo abundances $dn/dM$, 
which reflects $P(k)$ at scales too small to be directly measured in the angular power spectrum~\cite{Sekiguchi:2013lma}.
If there were no effects on the  shape of $C_l^{\rm (21cm)}$, the parameters would be indistinguishable. However, in reality, 
the response of the overall amplitude to the higher order running parameters is dependent on redshifts, which enables to disentangle
such parameters. Such redshift dependence can be seen by 
comparing panels of Fig.~\ref{fig:dcl21cm}, each of which has distinct $\Delta z=z-z'$.

\begin{table}
  \begin{center}
    \begin{tabular}{lcc}
      \hline\hline
      & SKA & FFTT \\
      \hline
      total effective area $A_{\rm tot}$ [m$^2$]& $10^5$ & $10^7$ \\
      bandwidth $\Delta \nu$ [MHz] & \multicolumn{2}{c}{$1$} \\
      beam width $\Delta \theta$ [arcmin] & \multicolumn{2}{c}{$9$} \\
      integration time $t$ [hour] & \multicolumn{2}{c}{$1000$} \\
      \hline\hline
    \end{tabular}
  \end{center}
  \caption{Specification of 21~cm surveys.}
  \label{tab:spec_21cm}
\end{table}
\begin{table}
  \begin{center}
    \begin{tabular}{lcccccccccc}
      \hline\hline
      &\multicolumn{3}{c}{Planck} &\quad\quad&\multicolumn{5}{c}{COrE} \\
      \hline
      band frequency [GHz] &100 & 147 & 217 &\quad\quad& 105 & 135 & 165 & 195 & 225 \\
      beam width $\Delta\theta$ [arcmin] & 9.9 & 7.2 & 4.9 &\quad\quad& 10.0 & 7.8 & 6.4 & 5.4 & 4.7 \\
      Temperature noise $\Delta_T$ [$\mu$K arcmin] & 31.3 & 20.1 & 28.5 &\quad\quad& 2.68 & 2.63 & 2.67 & 2.63 & 2.64 \\
      Polarization noise $\Delta_P$ [$\mu$K arcmin] & 44.2 & 33.3 & 49.4 &\quad\quad& 4.63 & 4.55 & 4.61 & 4.54 & 4.57 \\
      \hline\hline
    \end{tabular}
  \end{center}
  \caption{Specification of CMB surveys.}
  \label{tab:spec_cmb}
\end{table}

\begin{figure}
  \begin{center}
   \hspace{0mm}\scalebox{2.5}{\includegraphics{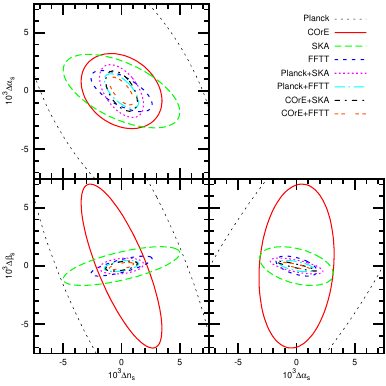}}
  \end{center}
  \vspace{20mm}
  \caption{Expected constraints on the primordial power spectrum 
  from observations of 21~cm signals from minihalos in combination
 with the CMB.
  Contours depict 1$\sigma$ errors with other cosmological parameters being marginalized.
  We assume 21~cm signals from minihalos can be measured down to $z_{\rm min}=6$.
  }
  \label{fig:zmin06}
\end{figure}

Fig.~\ref{fig:zmin06} shows constraints on the primordial power spectrum from 21~cm, CMB, and combinations of these observations.
In Table~\ref{tab:constraint} we summarize expected 1$\sigma$ constraints on $n_s$, $\alpha_s$ and $\beta_s$ from 
different combinations of observations. In the figure,  we have assumed 21~cm line observations can measure 
the signal from minihalos at a redshift between $z_{\rm min}=6$ and $z_{\rm max}=20$.

First of all, it is remarkable in the figure that 21~cm line power spectrum from minihalos~(i.e.~SKA and FFTT) is 
competitive or even more powerful in comparison with the CMB observations~(i.e.~Planck and COrE) as a probe of the primordial 
power spectrum. In particular, $C_l^{\rm (21cm)}$ can measure $\beta_s$ more tightly
than $C_l^{\rm (CMB)}$. This is because the abundance of minihalos is sensitive to the linear matter 
fluctuations at very small scales, which is difficult to be measured directly.
Meanwhile, $C_l^{\rm (21cm)}$ can also measure fluctuations at larger scales through the spectral 
shape, which constrains $n_s$ and $\alpha_s$.

Once observations of 21~cm line fluctuations from minihalos and the CMB are combined, we can probe matter 
fluctuations over a wide range of scales. Combinations of these two different observations offer a great advantage by the lever-arm effect
when we try to constrain the primordial power spectrum that is close to the power-law as in Eq.~\eqref{eq:pow}.
In other words, combinations of these observations can break parameter degeneracies that each observation suffers from by itself.
This is most notable in Fig.~\ref{fig:zmin06} when Planck and SKA are combined.

\begin{table}
  \begin{center}
    \begin{tabular}{lccc}
      \hline\hline
      & $10^{-3}\Delta n_s$ & $10^{-3}\Delta \alpha_s$ & $10^{-3}\Delta \beta_s$ \\
      \hline
	Planck & 7.7 & 10.7 & 15.1 \\
	COrE & 3.2 & 2.9 & 6.5 \\
	SKA & 4.6 & 2.9 & 1.5 \\
	FFTT & 2.4 & 1.6 & 0.79 \\
	Planck+SKA & 1.7 & 2.0 & 0.63 \\
	Planck+FFTT & 1.3 & 1.3 & 0.44 \\
	COrE+SKA & 1.2 & 1.6 & 0.39 \\
	COrE+FFTT & 0.95 & 1.1 & 0.28 \\
      \hline\hline
    \end{tabular}
  \end{center}
  \caption{Constraints on parameters for the primordial power spectrum. 
  For 21~cm observations, $z_{\rm min}=6$ is assumed.}
  \label{tab:constraint}
\end{table}

On the other hand, as  can be read off by Eq.~\eqref{eq:noise},
the signal-to-noise ratio in observations of the 21cm signal from minihalos
rapidly increases at low redshifts. Therefore, the resultant constraints 
are expected to be dependent of $z_{\rm min}$, the minimum redshift until when minihalos can be observed.
In order to examine the dependence, 
we have evaluated expected (marginalized) 1$\sigma$ uncertainties for the determination of 
$n_s$, $\alpha_s$ and $\beta_s$ from two datasets, Planck+SKA and COrE+FFTT 
with $z_{\rm min}$ being 4, 6 (baseline), 8 and 10, 
which are summarized in Table~\ref{tab:zmin}.
As expected, the constraints become less tight as $z_{\rm min}$ is increased
although the changes are not so significant.
For example, the change  of $z_{\rm min}$ from $6$ to $8$ 
degrades the constraint on each of the parameters by around 50\%
for Planck+SKA. On the other hand, for the combination of COrE+FFTT, the degradation becomes
modest; between $z_{\rm min}=6$ and 10, the constraint changes only by 10\%.

\begin{table}
  \begin{center}
    \begin{tabular}{llccc}
      \hline\hline
      & $z_{\rm min}$ & $10^{-3}\Delta n_s$ & $10^{-3}\Delta \alpha_s$ & $10^{-3}\Delta \beta_s$ \\
      \hline
	\multirow{4}{*}{Planck+SKA} 
	& 4 & 1.4 & 1.4 & 0.40 \\
	& 6 & 1.7 & 2.0 & 0.63 \\
	& 8 & 2.3 & 3.0 & 0.85 \\
	& 10 & 3.6 & 4.7 & 1.2 \\
	\hline
	\multirow{4}{*}{COrE+FFTT} 
	& 4 & 0.85 & 0.96 & 0.24 \\
	& 6 & 0.95 & 1.1 & 0.28 \\
	& 8 & 1.0 & 1.2 & 0.31 \\
	& 10 & 1.1 & 1.3 & 0.33 \\
      \hline\hline
    \end{tabular}
  \end{center}
  \caption{Dependence of the constraints on $z_{\rm min}$ for the data sets, Planck+SKA and COrE+FFTT.
  }
  \label{tab:zmin}
\end{table}

\subsection{Implication for the constraint on the inflationary models}
\label{sub_sec:models}

Now let us consider the implication of our results for the constraint on the inflationary models.
As shown in the Planck paper~\cite{Ade:2015lrj}, the parameter plane in
terms of the spectral index and the tensor-to-scalar ratio~($n_s$-$r$ plane)
is  useful to constrain the inflationary models (for the predictions of various inflation models, see \cite{Martin:2013tda}).
In fact, it implies that the simple chaotic inflationary models with the inflaton's potential $V\propto \phi^n$
are  almost ruled out for  $n\gtrsim2$
and the so-called $R^2$-inflation model~\cite{Starobinsky:1980te, Nariai:1971sv, Tomita:2016tcj} seems to be favored.
However, 
once the multi-field inflationary models are taken into account such
as the curvaton model \cite{Enqvist:2001zp,Lyth:2001nq,Moroi:2001ct}, modulated reheating model \cite{Dvali:2003em,Kofman:2003nx} and so on,
where a light scalar field other than inflaton exists and its fluctuations also contribute to primordial fluctuations,
not only $R^2$-inflation but also several
inflationary models are well inside the allowed region
on the $n_s$-$r$
plane~\cite{Langlois:2004nn,Lazarides:2004we,Moroi:2005kz,Moroi:2005np,Ichikawa:2008iq,Fonseca:2012cj,Enqvist:2013paa,Vennin:2015vfa,Ichikawa:2008ne,
Suyama:2010uj,Enqvist:2015njy,Fujita:2014iaa}.
It is well-known that one of the powerful tools to distinguish single-field inflationary models from the multi-field 
ones is the non-Gaussianity of the primordial fluctuations,
especially the so-called local-type non-Gaussianity.
In general, the single-field models predict  small local-type non-Gaussianity, while  multi-field models could generate relatively larger one.
However, at the level of current constraint on non-Gaussianity from Planck \cite{Ade:2015ava}, we cannot differentiate between single-field and multi-field models.

Here, we  discuss the potential of the runnings of the spectral index to distinguish among inflation models, 
having our Fisher analysis results given in the previous section in mind. 
In the following, we choose some representatives of single-field and multi-field models and apply the  expressions
for the spectral parameters provided in Appendix~\ref{app:higher_power_slow_roll}.
As examples for  single-field models, we consider the $R^2$-inflation and brane-inflation. 
For multi-field models, the natural-  and inverse monomial-spectator models are investigated. 
We take the model parameters in such a way that these models are consistent with the current observations 
and hardly distinguished from one another to date (see Section \ref{sec:pred}).

\subsubsection{$R^2$-inflation}

$R^2$-inflation was proposed in Refs.~\cite{Starobinsky:1980te, Nariai:1971sv, Tomita:2016tcj}, where the inflationary phase 
can be realized by the higher order curvature term, $R^2$. It has been known that this model corresponds to the single-field
 inflation in Einstein frame with the potential of
\begin{eqnarray}
V(\phi) = \Lambda^4 \left( 1 - e^{-\sqrt{2/3} \phi / M_{\rm pl}}\right)^2.
\end{eqnarray} 
Based on the slow-roll approximation, the $e$-folding number measured from the time when the pivot scale $k_0$ left the Hubble radius during the inflation
to the end of inflation can be estimated as
\begin{eqnarray}
N_0 = \int^{t_e}_{t_0} H dt \simeq - {1 \over M_{\rm pl}^2} \int^{\phi_e}_{\phi_0} {V \over V'} d\phi,
\end{eqnarray}
where the index $e$ and $0$ respectively represent the time when the inflation ends and the pivot scale $k_0$ left the Hubble radius during the inflation.
For the $R^2$-inflation model, this $e$-folding number can be obtained as
\begin{eqnarray}
N_0 \simeq {3 \over 4} e^{\sqrt{2/3} \phi_0 / M_{\rm pl}},
\end{eqnarray}
where we keep only the leading term.
The slow-roll parameters are also given in terms of $N_0$ as
\begin{eqnarray}
\epsilon \simeq {3 \over 4} {1 \over N_0^2},
\qquad
\eta \simeq - {1 \over N_0},
\qquad
\xi^{(2)} \simeq {1 \over N_0^2},
\qquad
\sigma^{(3)} \simeq - {1 \over N_0^3},
~ \cdots ,
\end{eqnarray}
where we have assumed $N_0 \gg 1$.
For the single-field models, since the tensor-to-scalar ratio and the spectral index are respectively given by
\begin{eqnarray}
n_s - 1 = - 6 \epsilon + 2 \eta,
\qquad
r = 16 \epsilon,
\end{eqnarray}
we can estimate  $n_s$ and $r$ from the $e$-folding number $N_0$
for the $R^2$-inflation as
\begin{eqnarray}
n_s - 1 \simeq - {2 \over N_0} \left( \simeq -(3.3~ {\text{--}}~ 4)  \times 10^{-2} \right) , 
\qquad
r = {12 \over N_0^2}  \left( \simeq  (3.3~  {\text{--}} ~4.8) \times 10^{-3} \right) \quad ({\rm for}~N_0 = 50 ~{\text{--}} ~ 60), \notag \\
\end{eqnarray}
at the leading order in $N_0$.

\begin{figure}[htbp]
\begin{center}
\includegraphics[width=150mm]{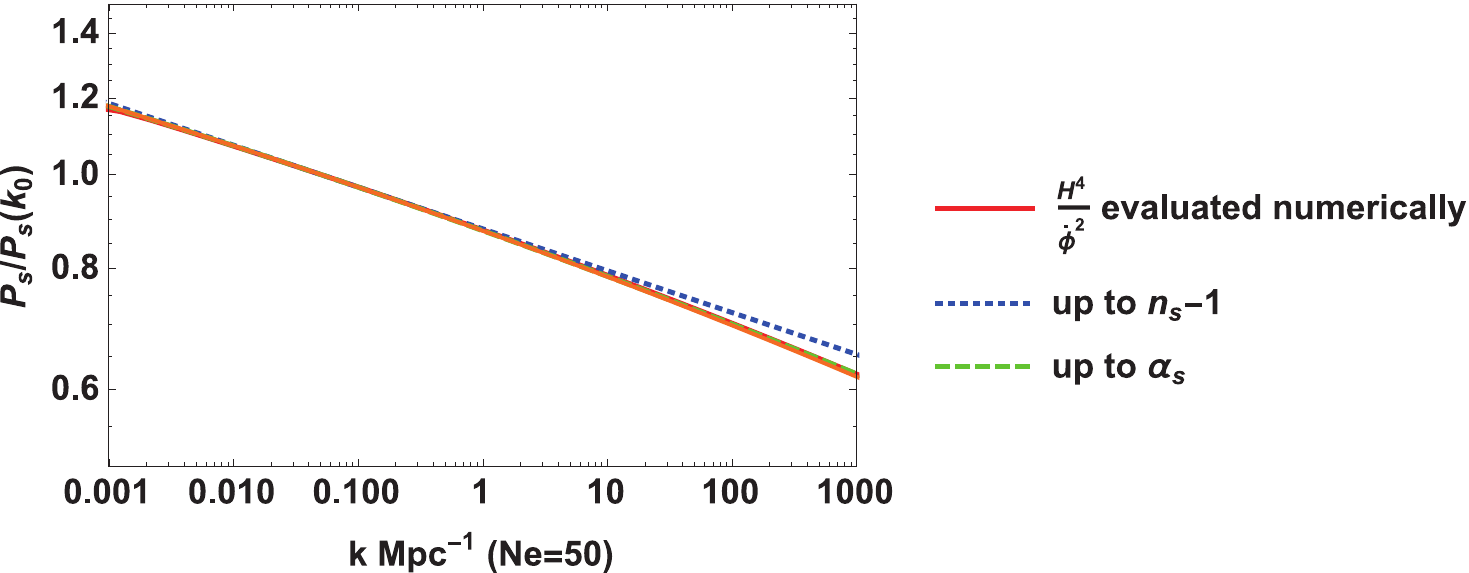}
\end{center}
\caption{The power spectrum of the primordial curvature perturbations
 generated from the $R^2$-inflation. The red line is obtained by numerically calculating the background dynamics and evaluating 
$H^4/\dot{\phi}_0^2$. The blue dotted line corresponds to the one calculated from the power-law form 
given by \eqref{eq:pow2} including up to $n_s$, and the green dashed line corresponds to that up to $\alpha_s$.}
\label{fig:ipower}
\end{figure}

The scales of the primordial fluctuations on which we focus  here are around $20~{\rm Mpc}^{-1} < k < 500~{\rm Mpc}^{-1}$ as we have discussed.
In Fig. \ref{fig:ipower}, we plot the power spectrum of the primordial curvature perturbations generated from $R^2$-inflation, numerically calculated one 
and approximated by the power-law expansion form. 
From this figure, one can find that at minihalos' scales the power spectrum only including  $n_s $ deviates from the numerically-evaluated power spectrum.
On the other hand, by using the expression for the power spectrum up to $\alpha_s$ seems to be in good agreement with the numerical result.
Hence   we need to take account of a higher order running to test 
inflationary models using observations of small scales such as minihalos.
Based on Appendix~\ref{app:higher_power_slow_roll}, $\alpha_s$ and $\beta_s$ in the $R^2$-inflation can be respectively obtained as
\begin{eqnarray}
\alpha_s \simeq - {2 \over N_0^2} \left( \simeq - (5.5 ~\text{--}~ 8)  \times 10^{-4} \right), 
\qquad
\beta_s \simeq - {4 \over N_0^3}  \left( \simeq - ( 1.9 ~\text{--}~ 3.2)\times 10^{-5} \right)
\quad ({\rm for}~N_0 = 50 ~\text{--}~  60), \notag \\
\end{eqnarray}
at the leading order in $N_0$.

\subsubsection{Brane  inflation}
The potential of the brane inflation models is phenomenologically given as (see \cite{Lorenz:2007ze,Martin:2013tda} and references therein)
\begin{eqnarray}
V(\phi) = V_0 \left[ 1 - \left( \frac{\phi}{\mu} \right)^{-p} \right],
\end{eqnarray}
where $V_0$ represents the energy scale of the model and  $p$ and $\mu$ are also model parameters.

The slow-roll parameters in this model are given as
\begin{eqnarray}
\label{eq:SR_BI}
\begin{array}{lll}
&&\epsilon =  \displaystyle\frac{p^2}{ 2\tilde{\mu}^2 \tilde{\phi}_0^2 (\tilde{\phi}_0^p -1)^2},~~~~\eta =  - \displaystyle\frac{p (p+1)}{ \tilde{\mu}^2 \tilde{\phi}_0^2 (\tilde{\phi}_0^p -1)}, \cr\cr
&&\xi^{(2)} =   \displaystyle\frac{p^2 (p+1) (p+2)}{ \tilde{\mu}^4 \tilde{\phi}_0^4 (\tilde{\phi}_0^p -1)^2}, 
~~~\sigma^{(3)} =  - \displaystyle\frac{p^3 (p+1) (p+2) (p+3)}{ \tilde{\mu}^6 \tilde{\phi}_0^6 (\tilde{\phi}_0^p -1)^3},
\end{array}
\end{eqnarray}
where we have defined $\tilde{\mu}$ and $\tilde{\phi}_0$ as $\tilde{\mu} \equiv \mu / M_{\rm pl}$ and $\tilde{\phi}_0 \equiv \phi_0 / \mu$. 
The number of $e$-folds can be expressed as 
\begin{equation}
\label{eq:Ne_BI}
N_0 = \frac{\tilde{\mu}^2}{2p} \left[  \frac{2}{p+2} \left( \tilde{\phi}_0^{p+2} -  \tilde{\phi}_e^{p+2} \right)  -  \left( \tilde{\phi}_0^{2} -  \tilde{\phi}_e^{2} \right) \right],
\end{equation}
with $\phi_e$ being the value of $\phi$ at the end of inflation, at which one of the slow-roll parameters exceeds unity.
Assuming that $\tilde{\phi}_0 \gg 1$ and $\tilde{\phi}_0 \gg \tilde\phi_e$ in Eq.~\eqref{eq:Ne_BI}, $\tilde{\phi}_0$ can be  approximated by 
\begin{equation}
\tilde{\phi}_0  \simeq \left[  p(p+2) \frac{N_e}{\tilde{\mu}^2} \right]^{1/(p+2)}.
\end{equation}
The spectral index $n_s$, the tensor-to-scalar ratio $r$, the running parameters $\alpha_s$ and $\beta_s$ can be calculated in the same way by putting 
the slow-roll parameters given in Eq.~\eqref{eq:SR_BI} into 
the formulas Eqs.~\eqref{eq:alpha_phi} and \eqref{eq:beta_phi}.

\subsubsection{Natural-spectator model}

Next we consider a multi-field model. To predict the spectral index, its higher order runnings and the tensor-to-scalar ratio, 
we do not have to specify the spectator field model itself, but need to specify the potential for the spectator field $\chi$. 
Here we take a quadratic potential for $\chi$ as $V(\chi) = \frac12 m_\chi^2 \chi^2$ with $m_\chi$ being the mass of the spectator field and 
assume that $m_\chi$ is much smaller than the Hubble parameter during inflation, 
which gives $\eta_\chi \simeq 0$ and $\xi_\chi^{(2)} = \sigma_\chi^{(3)} = 0$.
Furthermore, we also have to specify the inflaton potential.

A simplest natural inflation model is characterized by the potential~\cite{Freese:1990rb, Adams:1992bn}:
\begin{eqnarray}
V(\phi) = \Lambda^4 \left[ 1 - \cos \left( \phi \over f \right) \right] \quad ( 0 \leq \phi \leq \pi f),
\end{eqnarray}
where $\Lambda$ denotes the energy scale of the model
and $f$ corresponds to some breaking scale which  determines the curvature of the potential.
Based on the slow-roll approximation, the $e$-folding number for the natural inflation is expressed as
\begin{eqnarray}
N_0 = {f^2 \over M_{\rm pl}^2} \ln \left| {\cos^2 (\phi_e/2f) \over \cos^2 (\phi_0/2f) }\right|,
\end{eqnarray}
where $\phi_e$ denotes the field value at the end of inflation.
The slow-roll parameters are respectively given by
\begin{eqnarray}
\label{eq:slow_roll_natural}
\begin{array}{lll}
&&\epsilon = \displaystyle{M_{\rm pl}^2 \over 2 f^2} {\cos^2 (\phi_0/2f) \over \sin^2 (\phi_0/2f)},~~~\eta = {M_{\rm pl}^2 \over f^2} {1 - 2 \sin^2 (\phi_0/2f) \over 2 \sin^2 (\phi_0/2f)}, \cr\cr
&&\xi^{(2)} = - \displaystyle{M_{\rm pl}^4 \over f^4}{\cos^2 (\phi_0/2f) \over \sin^2 (\phi_0/2f)},
~~~\sigma^{(3)} = - {M_{\rm pl}^6 \over  f^6} {\cos^2 (\phi_0/2f) (1 - 2 \sin^2 (\phi_0/2f)) \over  2 \sin^2 (\phi_0/2f) } .
\end{array}
\end{eqnarray}
The end of inflation is defined by $\epsilon = 1$, and it gives
\begin{eqnarray}
\cos^2 {\phi_e \over 2f} = {2f^2/M_{\rm pl}^2 \over 1 + 2 f^2/M_{\rm pl}^2}.
\end{eqnarray}
The prediction for the spectral index, its higher order runnings and the tensor-to-scalar ratio can be calculated by putting the slow-roll parameters Eqs.~\eqref{eq:slow_roll_natural}
into the formulas \eqref{eq:ns_eff}, \eqref{eq:alpha_eff}, \eqref{eq:beta_eff} and \eqref{eq:r_multi} given in Appendix~\ref{app:higher_power_slow_roll}.

\subsubsection{Inverse monomial-spectator model }
The potential of the inverse monomial inflation model is given by 
\begin{equation}
V (\phi) = V_0 \left( \frac{\phi}{M_{\rm pl}} \right)^{-p},
\end{equation}
where $V_0$ represents the energy scale of the model and $p$ is assumed to be positive.
In the slow-roll approximation, the number of $e$-folds is written as
\begin{equation}
\label{eq:Ne_IMI}
N_0 = \frac{1}{2p} \left[  \left( \frac{\phi_e}{M_{\rm pl}} \right)^2 - \left( \frac{\phi_0}{M_{\rm pl}} \right)^2 \right].
\end{equation}
Notice that the inflation does not end by the slow-roll violation in this model, hence it needs some mechanism to exit from the inflationary era.
Here we just assume some mechanism works to stop the inflation. 
The slow-roll parameters are given by
\begin{eqnarray}
\begin{array}{lll}
&\epsilon = \displaystyle\frac12 p^2 \left( \frac{\phi_0}{M_{\rm pl}}  \right)^{-2},
&~\eta =  p (p+1) \left( \displaystyle\frac{\phi_0}{M_{\rm pl}} \right)^{-2}, \\
&\xi^{(2)} =   p^2 (p+1) (p+2)\left( \displaystyle\frac{\phi_0}{M_{\rm pl}} \right)^{-4}, 
&~\sigma^{(3)} =  p^3 (p+1) (p+2) (p+3) \left( \displaystyle\frac{\phi_0}{M_{\rm pl}} \right)^{-6}.
\end{array}
\end{eqnarray}
From Eq.~\eqref{eq:Ne_IMI}, one can express $\phi_0$ as 
\begin{equation}
\label{eq:phi_N}
 \left( \frac{\phi_0}{M_{\rm pl}} \right)^2 = \left( \frac{\phi_e}{M_{\rm pl}} \right)^2 - 2p N_0.
\end{equation}
As mentioned above, the inflation should end by some mechanism (not by the slow-roll violation), 
$\phi_e$ depends on the mechanism. Since varying $\phi_e$ corresponds to changing $\phi_0$ as seen from
Eq.~\eqref{eq:phi_N},  we take $\phi_0$ as a phenomenological parameter which 
is chosen to obtain $n_s$ and $r$ in accordance with observational constraint.

\subsubsection{Predictions of the models} \label{sec:pred}
\begin{figure}
  \begin{center}
    \begin{tabular}{cc}
    \hspace{-0mm}\scalebox{.4}{\includegraphics{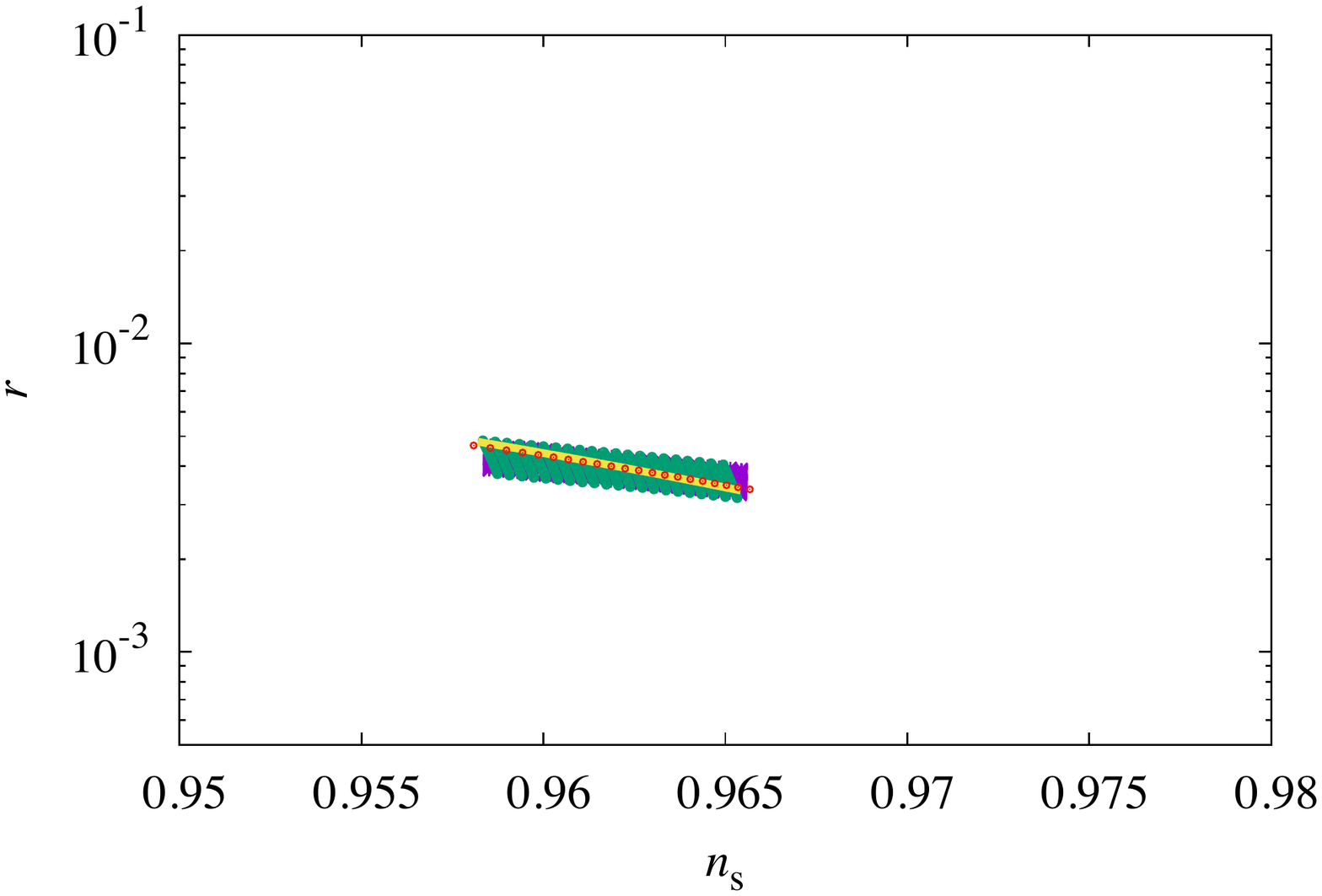}} &
    \hspace{-0mm}\scalebox{.4}{\includegraphics{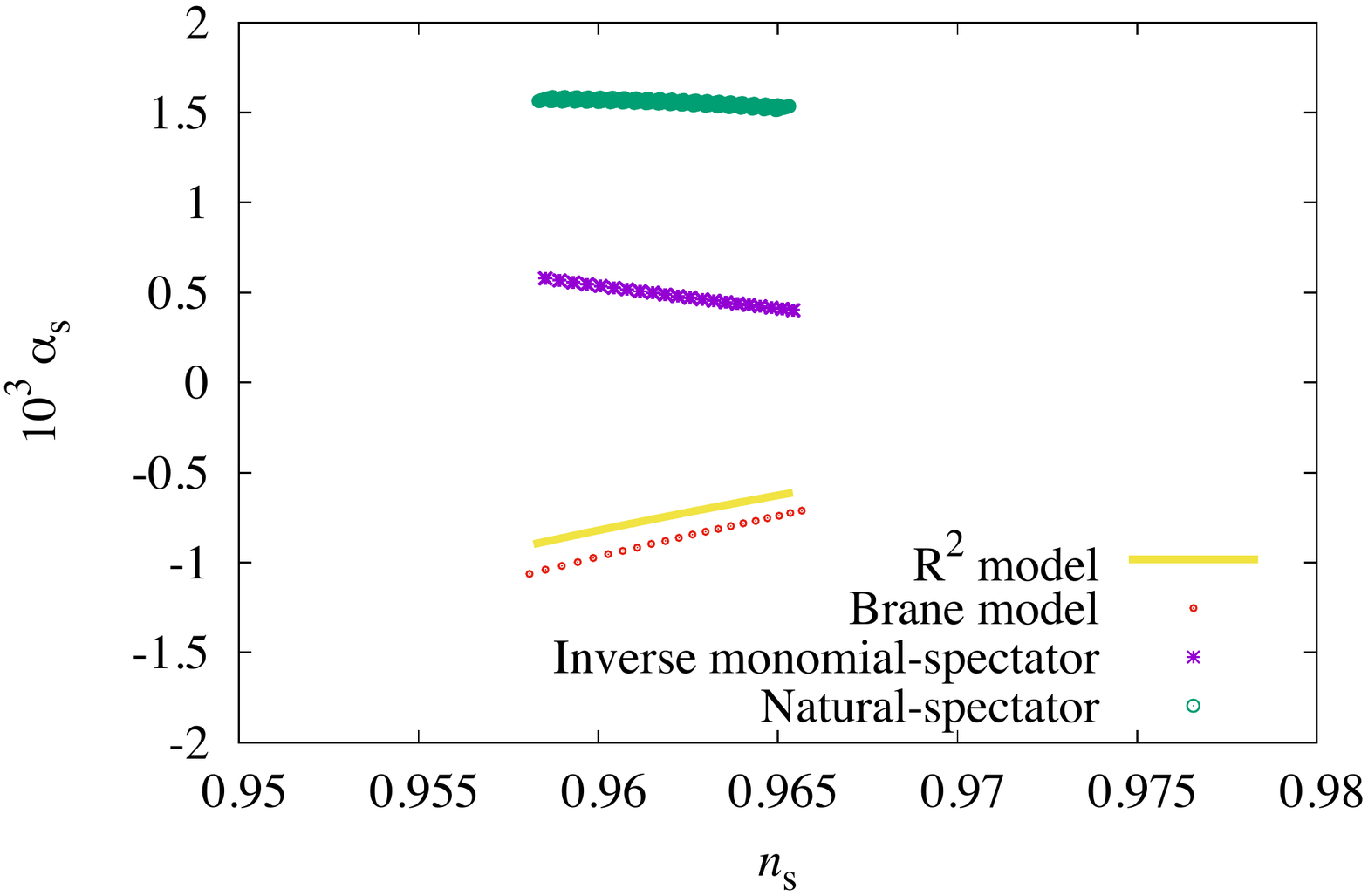}} \vspace{-60mm}\\
    \hspace{-0mm}\scalebox{.4}{\includegraphics{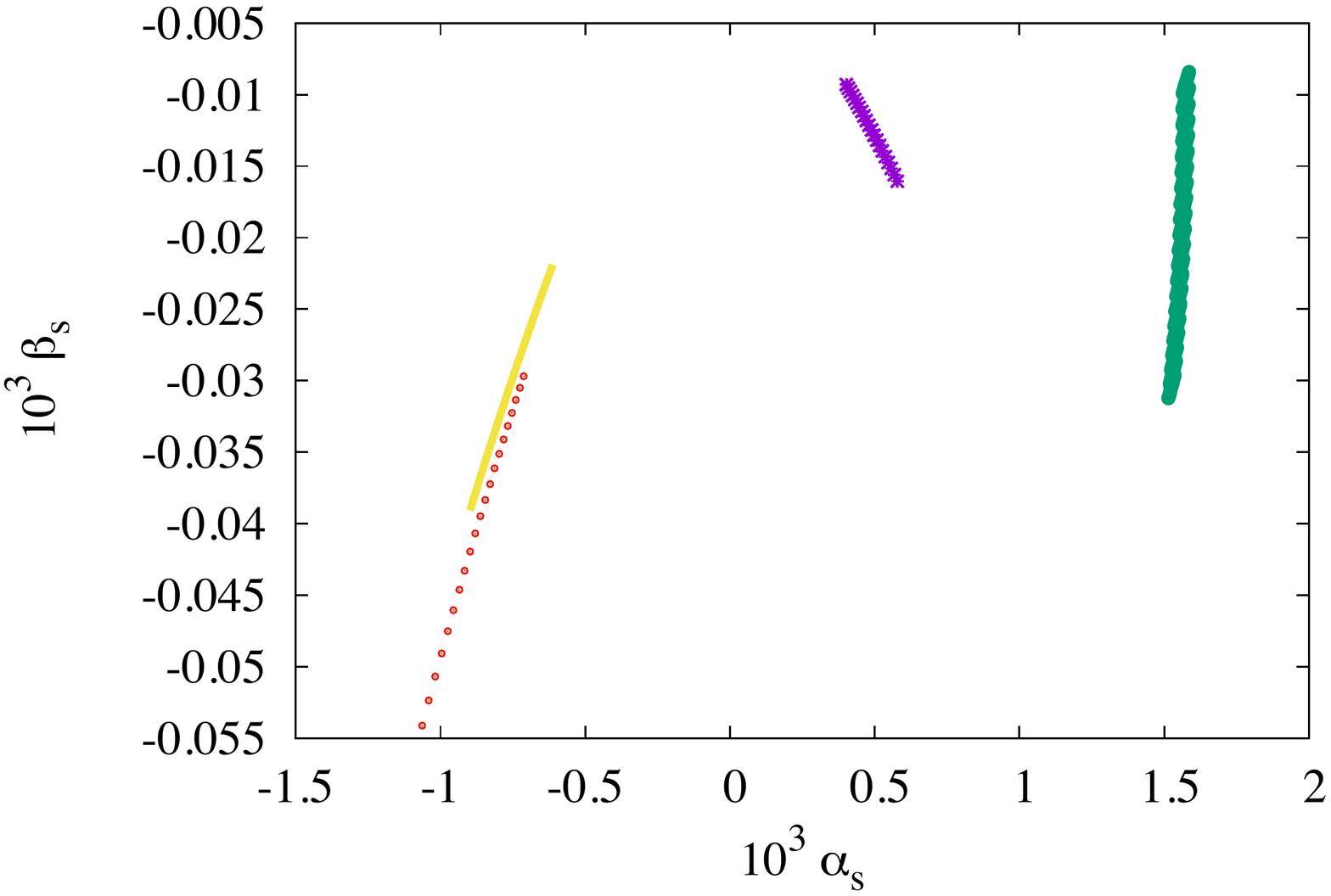}} & 
    \end{tabular}
  \end{center}
  \vspace{-30mm}
  \caption{Predictions for $n_s$, $\alpha_s$, $\beta_s$ and $r$
  for the  $R^2$-inflation~(yellow), brane inflation~(orange),  the natural-spectator model~(green) and the inverse monomial-spectator model~(purple).
 The upper left panel shows the predictions in the $n_s$ -- $r$ plane,
and the upper right one shows those in the $n_s$ -- $\alpha_s$ plane. The lower panel
shows the predictions in the $\alpha_s$ -- $\beta_s$ plane.
    }
  \label{fig:prediction}
\end{figure}

In Fig.~\ref{fig:prediction}, 
we plot the predictions for $r$, $n_s$, $\alpha_s$ and $\beta_s$ for the
$R^2$-inflation (yellow), brane-inflation (magenta), 
natural-spectator (green) and inverse monomial-spectator (purple) models.
The upper left, upper right and  lower panels  show the predictions in the $n_s$ -- $r$, $n_s$ -- $\alpha_s$ and $\alpha_s$ -- $\beta_s$ planes, respectively.

For the $R^2$ model, we take $50  < N_0 < 60$, which gives a very good fit to the current Planck observations \cite{Ade:2015lrj}.
For other models, we choose the values of the parameters such that the prediction for  $n_s$ and $r$ become almost the same with those for $R^2$ inflation,  
which can be seen in the upper left panel of Fig.~\ref{fig:prediction}.
Also some of the parameters are chosen to give a correct amplitude for the primordial power spectrum.
For the brane inflation, we consider the case with $p=4$ and  $\mu= 2.6 M_{\rm pl}$,  and assume $ 43 < N_0 < 52 $.
For the natural-spectator model, we take $f = 3.5 M_{\rm pl}$ and $51 < N_0 < 54$ for the inflaton sector and assume 
the fractional contribution of the spectator, $Q_\chi$ (defined in Eq.~\eqref{eq:Qchi}),  to be $0.5 < Q_\chi < 0.6$.
For the inverse monomial-spectator model, we consider the case with $p=6$ and take $Q_\chi \simeq 0.98$ (or precisely speaking, 
by using $R$ defined in Eq.~\eqref{eq:def_R}, we take $ 75 < R  <80$.)
The field value at the end of inflation $\phi_e$ is varied to be tuned to give almost the degenerate prediction for $n_s$ and $r$ with $R^2$ inflation 
as depicted in the upper left panel of Fig.~\ref{fig:prediction}.

As seen in the upper left in Fig.~\ref{fig:prediction}, models discussed here can give almost degenerate predictions for $n_s$ and $r$, 
however, when we compare the predictions of these models in the $n_s$ -- $\alpha_s$ and the $\alpha_s$ -- $\beta_s$ planes, 
we can see that those models give different runnings, which would be helpful to distinguish the model.
It should be noted here that some models could be easily differentiated by using  the expected constraint from future mininalo observations 
on the $n_s$ -- $\alpha_s$ and $\alpha_s$ -- $\beta_s$ planes  shown in  Fig.~\ref{fig:zmin06}.  On the other hand, some models are still difficult to be
distinguished from the expected constraint investigated in this paper.

\section{Conclusion}
\label{sec:conclusion}
In this paper, we have developed the formalism to use the angular power spectrum of 21~cm line fluctuations from minihalos
to constrain cosmological models, instead of adopting 
the analysis based on the pixel variance, which have been used in  previous studies.
Our formulation can take account of cross-correlations not only in angular domains, but also in redshift ones.
The advantage in considering such cross-correlations over the simple pixel variance is that
we can decode the scale-dependences on top of the increase in the statistics.

By making use of the formalism, we have forecasted constraints on the power spectrum of primordial perturbations
from future observations of  21~cm line fluctuations from minihalos in
combination with the CMB. In particular, we have focused on the spectral index $n_s$ and its runnings $\alpha_s$ and $\beta_s$.
Our results exhibit the potential of the angular power spectrum of 21~cm line fluctuations 
as a promising probe of primordial power spectrum. 
Particularly, the synergy in combination with the CMB is remarkable due to the lever-arm effect 
since 21~cm fluctuations from minihalos are sensitive to those on small scales, while CMB observations 
probe those on large scales. 
For any combinations of CMB and 21~cm observations (Planck+SKA, Planck+FFTT, COrE+SKA, COrE+FFTT), 
the spectral runnings can be probed down to the level of $\alpha_s \sim {\cal O}(10^{-3})$ and $\beta_s \sim {\cal O}(10^{-4})$ (see Table~\ref{tab:constraint}).

We have also discussed  implications of future constraints on the runnings for differentiating inflationary models. 
For the purpose, we considered several representative models of single- and multi-field models including $R^2$-inflation.
We have shown that future sensitivities of 21~cm signals from minihalos on the runnings $\alpha_s$ and $\beta_s$ would be helpful to differentiate inflationary models in some 
cases. Although we have just looked at some representative models, a more exhaustive study of inflationary models from the viewpoint of 
the spectral runnings would give more insight to really understand the mechanism of inflation.


\bigskip
\section*{Acknowledgments}

TT thanks Sami Nurmi for helpful discussions.
This work is partially supported by JSPS KAKENHI Grant Number
15K05084 (TT), 15K17646 (HT), 15K17659 (SY),  MEXT KAKENHI Grant Number 15H05888 (TT, SY), 16H01103 (SY),
and IBS under the project code, IBS-R018-D1 (TS).

\appendix
\bigskip
\bigskip
\noindent 
{\Large \bf Appendix}

\section{Expressions for higher order runnings with slow-roll parameters}
\label{app:higher_power_slow_roll}

In Eq.~\eqref{eq:pow}, we have expanded the primordial power spectrum ${\cal P}_s$ up to the 2nd  order. 
However, in principle, the primordial power spectrum can be expanded up to arbitrarily higher orders. Here we truncate the expansion at 
the 4th  order, in which  ${\cal P}_s$ is given by
\begin{equation}
\mathcal P_s(k)=A_s\left(\frac{k}{k_0}\right)^{\displaystyle n_s-1+\frac12 \alpha_s \ln\left( k/k_0 \right)+\frac{1}{3!} \beta_s \ln^2 \left(k/ k_0 \right) 
+ \frac{1}{4!} \gamma_s \ln^3 \left(k/ k_0 \right) + \frac{1}{5!} \delta_s \ln^4 \left(k/ k_0 \right) },
\label{eq:pow2}
\end{equation}
where the  runnings $\alpha_s, \beta_s, \gamma_s$ and $\delta_s$ are defined as 
\begin{equation}
\begin{array}{ll}
\alpha_s \equiv \displaystyle\frac{d^2 \ln \mathcal P_s (k)}{(d\ln k)^2} \Biggr|_{k = k_0}, &
\qquad
\beta_s \equiv \displaystyle\frac{d^3 \ln \mathcal P_s (k)}{(d\ln k)^3}\Biggr|_{k = k_0},  \\  \\
\gamma_s \equiv \displaystyle\frac{d^4 \ln \mathcal P_s (k)}{(d\ln k)^4} \Biggr|_{k = k_0}, &
\qquad
\delta_s \equiv \displaystyle\frac{d^5 \ln \mathcal P_s (k)}{(d\ln k)^5} \Biggr|_{k = k_0}.
\end{array}
\end{equation}
Below we give explicit expressions for these runnings  using the slow-roll parameters 
for the single-field and multi-field models. 

\subsection{Single-field case}
Assuming a slow-roll single-field inflation model with a canonical kinetic term, 
$n_s$ and the running parameters can be explicitly written down with the slow-roll parameters, which are defined using the inflaton potential $V(\phi)$ as 
\begin{equation}
\label{eq:SR_phi}
\begin{array}{llll}
&  \epsilon \equiv \displaystyle\frac12 M_{\rm pl}^2 \left( \frac{V'}{V} \right)^2,
\qquad
& \eta \equiv  M_{\rm pl}^2  \displaystyle\frac{V^{\prime\prime}}{V},
\qquad
& \xi^{(2)} \equiv  M_{\rm pl}^4  \displaystyle\frac{V^{\prime}V^{\prime\prime\prime}}{V^2}, \\
&\sigma^{(3)} \equiv  M_{\rm pl}^6  \displaystyle\frac{(V^{\prime})^2 V^{(4)}}{V^3},
\qquad
& \tau^{(4)} \equiv  M_{\rm pl}^8  \displaystyle\frac{(V^{\prime})^3 V^{(5)}}{V^4},
\qquad
& \zeta^{(5)} \equiv  M_{\rm pl}^{10}  \displaystyle\frac{(V^{\prime})^4 V^{(6)}}{V^5},
\end{array}
\end{equation}
where a prime denotes the derivative with respect to $\phi$.
Using these slow-roll parameters, the spectral index and the runnings are given by:
\begin{eqnarray}
\label{eq:ns_phi}
n_s -1 = &&-6 \epsilon + 2 \eta,   \\
\label{eq:alpha_phi}
\alpha_s = && -24\epsilon^2  + 16 \epsilon \eta  - 2 \xi^{(2)}, \\ 
\label{eq:beta_phi}
\beta_s =  && -192 \epsilon^3 + 192 \epsilon^2 \eta - 32 \epsilon \eta^2 - 24 \epsilon \xi^{(2)} + 2 \eta \xi^{(2)} + 2 \sigma^{(3)}, \\
\gamma_s = 
&& 
- 2304 \epsilon^4 
+ 3072 \epsilon^3 \eta 
- 1024 \epsilon^2 \eta^2 
- 384 \epsilon^2 \xi^{(2)}
+ 64 \epsilon \eta^3   
+ 148 \epsilon \eta \xi^{(2)}  
+36 \epsilon \sigma^{(3)}   \notag \\
&&  
 - 6 \eta \sigma^{(3)}
-2 \eta^2 \xi^{(2)} 
- 2  (\xi^{(2)})^2  
- 2 \tau^{(4)}, \\
\label{eq:delta_phi}
\delta_s  = && 
- 36864 \epsilon^5 
+ 61440 \epsilon^4 \eta 
- 30720 \epsilon^3 \eta^2 
- 7680 \epsilon^3 \xi^{(2)} 
+ 4736 \epsilon^2 \eta^3 
+ 5448 \epsilon^2 \eta \xi^{(2)}  \notag \\
&&  
+ 744 \epsilon^2 \sigma^{(3)} 
- 128 \epsilon \eta^4 
- 652 \epsilon \eta^2 \xi^{(2)} 
- 164 \epsilon (\xi^{(2)})^2
- 340 \epsilon \eta \sigma^{(3)} 
- 52 \epsilon \tau^{(4)}     \notag \\
&&
+ 2 \eta^3 \xi^{(2)}
+ 14 \eta^2 \sigma^{(3)}  
+ 8  \eta (\xi^{(2)})^2 
+ 12 \eta \tau^{(4)} 
+ 10 \xi^{(2)} \sigma^{(3)}   
+ 2 \zeta^{(5)}.
\label{eq:ns_runnings_slow_roll}
\end{eqnarray}

The tensor-to-scalar ratio is given by 
\begin{equation}
r = 16 \epsilon.
\end{equation}

\subsection{Multi-field case}

When a light scalar field $\chi$ other than the inflaton $\phi$ exists during the inflationary epoch\footnote{
Such another (light) scalar field is sometimes called a spectator field since it does not affect the inflationary dynamics. 
}, 
such a scalar field can also acquire primordial fluctuations and affect the present-day density fluctuations 
as in the curvaton model \cite{Enqvist:2001zp,Lyth:2001nq,Moroi:2001ct},
the modulated reheating model \cite{Dvali:2003em,Kofman:2003nx} and so on.
In general, fluctuations from the inflaton can also contribute to the primordial fluctuations, and therefore the power spectrum can be
given by the sum of those contributions\footnote{
This kind of model is called mixed inflaton and spectator field models, 
which has been studied in the context of the curvaton and modulated rehearing models 
\cite{Langlois:2004nn,Lazarides:2004we,Moroi:2005kz,Moroi:2005np,Ichikawa:2008iq,Fonseca:2012cj,Enqvist:2013paa,Vennin:2015vfa,Ichikawa:2008ne,Suyama:2010uj,Fujita:2014iaa,Enqvist:2015njy}.
}:
\begin{equation}
{\cal P}_s (k) = {\cal P}_s^{(\phi)} (k) + {\cal P}_s^{(\chi)} (k),
\end{equation}
where $ {\cal P}_s^{(\phi)}(k)$ and $ {\cal P}_s^{(\chi)} (k) $ are the primordial power spectra generated by 
the inflaton $\phi$ and another spectator field $\chi$. 
Due to the fact that the energy density of $\chi$  is subdominant during inflation, the spectral index and its runnings for $ {\cal P}_s^{(\chi)}$ are 
different from those for $ {\cal P}_s^{(\phi)}$ which are given in Eqs.~\eqref{eq:ns_phi}--\eqref{eq:delta_phi}.
To write down the spectral index and its runnings for  $ {\cal P}_s^{(\chi)}$, we also need to define the slow-roll parameters for $\chi$:
\begin{equation}
\begin{array}{llll}
& \eta_\chi \equiv   \displaystyle\frac{U^{\prime\prime}}{3H_\ast^2},
\qquad
& \xi_\chi^{(2)} \equiv    \displaystyle\frac{U^{\prime}U^{\prime\prime\prime}}{(3H_\ast^2)^2},
\qquad
&\sigma_\chi^{(3)} \equiv    \displaystyle\frac{(U^{\prime})^2 U^{(4)}}{(3H_\ast^2)^3}, \\
& \tau_\chi^{(4)} \equiv    \displaystyle\frac{(U^{\prime})^3 U^{(5)}}{(3H_\ast^2)^4},
\qquad
& \zeta_\chi^{(5)} \equiv    \displaystyle\frac{(U^{\prime})^4 U^{(6)}}{(3H_\ast^2)^5},
\end{array}
\end{equation}
where $U$ is a potential for $\chi$ field and a prime indicates the derivative 
with respect to the $\chi$ field. A superscript $(i)$ denotes the $i$-th derivative.
$H_\ast$ is the Hubble parameter at the horizon exit during inflation.
By using these slow-roll parameters along with those defined for $\phi$ provided in Eqs.~\eqref{eq:SR_phi},
the spectral index and its runnings for $\chi$ are given as\footnote{
Here we assume that there is no coupling between the inflaton and the spectator fields.
}
\begin{eqnarray}
n_s^{(\chi)} -1 = && -2 \epsilon + 2 \eta_\chi,   \\
\label{eq:alpha_chi}
\alpha_s^{(\chi)} = && 
 -8 \epsilon^2  + 4 \epsilon \eta + 4 \epsilon \eta_\chi - 2 \xi_\chi^{(2)}, \\
\beta_s^{(\chi)} =  && 
-64 \epsilon^3 
+ 56 \epsilon^2 \eta 
+ 24 \epsilon^2 \eta_\chi 
-  8 \epsilon \eta^2 
- 8 \epsilon \eta \eta_\chi 
 - 4 \epsilon \xi^{(2)}  
- 12 \epsilon \xi_\chi^{(2)} 
+  2 \eta_\chi \xi_\chi^{(2)} 
+ 2 \sigma_\chi^{(3)}, \notag \\ \\
\gamma_s^{(\chi)} = 
&& 
-768 \epsilon^4 
+ 944 \epsilon^3 \eta
 + 240 \epsilon^3 \eta_\chi
 -  288 \epsilon^2 \eta^2 
   - 160 \epsilon^2 \eta \eta_\chi 
     - 88 \epsilon^2 \xi^{(2)} 
       -  120 \epsilon^2 \xi_\chi^{(2)} \notag \\
&&       
 + 16 \epsilon \eta^3 
  + 16 \epsilon \eta^2 \eta_\chi 
  +  28 \epsilon \eta \xi^{(2)} 
  +  8 \epsilon \eta_\chi \xi^{(2)} 
  +  32 \epsilon \eta \xi_\chi^{(2)} 
  +  24 \epsilon \eta_\chi \xi_\chi^{(2)} 
    + 4 \epsilon \sigma^{(3)} 
\notag \\
&&       
  + 24 \epsilon \sigma_\chi^{(3)}  
  -  2 \eta_\chi^2 \xi_\chi^{(2)} 
    -  6 \eta_\chi \sigma_\chi^{(3)} 
  - 2 (\xi_\chi^{(2)})^2 
  - 2 \tau_\chi^{(4)}, \\
\delta_s^{(\chi)}  = 
&& 
-12288 \epsilon^5 
+  19360 \epsilon^4 \eta 
+  3360 \epsilon^4 \eta_\chi 
- 9120 \epsilon^3 \eta^2 
-  3360 \epsilon^3 \eta \eta_\chi 
-  2000 \epsilon^3 \xi^{(2)} 
-  1680 \epsilon^3 \xi_\chi^{(2)} 
\notag \\
&&       
+  1312 \epsilon^2 \eta^3 
+  800 \epsilon^2 \eta^2 \eta_\chi 
+  1296 \epsilon^2 \eta \xi^{(2)} 
+  240 \epsilon^2 \eta_\chi \xi^{(2)} 
+  960 \epsilon^2 \eta \xi_\chi^{(2)} 
+  360 \epsilon^2 \eta_\chi \xi_\chi^{(2)} 
 \notag \\
&&       
 +  128 \epsilon^2 \sigma^{(3)} 
   +  360 \epsilon^2 \sigma_\chi^{(3)}
- 32 \epsilon \eta^4 
-  32 \epsilon \eta^3 \eta_\chi 
-  132 \epsilon \eta^2 \xi^{(2)} 
-  56 \epsilon \eta \eta_\chi \xi^{(2)} 
-  28 \epsilon (\xi^{(2)})^2 
 \notag \\
&&       
-  80 \epsilon \eta^2 \xi_\chi^{(2)} 
-  80 \epsilon \eta \eta_\chi \xi_\chi^{(2)} 
-  40 \epsilon \eta_\chi^2 \xi_\chi^{(2)} 
-  40 \epsilon \xi^{(2)} \xi_\chi^{(2)}
 -  40 \epsilon (\xi_\chi^{(2)})^2 
 -  44 \epsilon \eta \sigma^{(3)}
  -  8 \epsilon \eta_\chi \sigma^{(3)} 
   \notag \\
&&       
   -  80 \epsilon \eta \sigma_\chi^{(3)}
    -  120 \epsilon \eta_\chi \sigma_\chi^{(3)} 
        - 4 \epsilon \tau^{(4)} 
    -  40 \epsilon \tau_\chi^{(4)} 
    +  2 \eta_\chi^3 \xi_\chi^{(2)} 
    +  14 \eta_\chi^2 \sigma_\chi^{(3)} 
     +  8 \eta_\chi (\xi_\chi^{(2)})^2 
      \notag \\
&&       
   +  12 \eta_\chi \tau_\chi^{(4)}
       +  10 \xi_\chi^{(2)} \sigma_\chi^{(3)} 
    + 2 \zeta_\chi^{(5)}.
\end{eqnarray}

Since ${\cal P}_s^{(\phi)}$ and ${\cal P}_s^{(\chi)}$  have different scale dependences, they should be treated separately.
However, we can define the effective spectral index and its runnings by using the total power spectrum as 
\begin{equation}
n_s^{\rm (eff)} - 1 =  \frac{ d \ln ( {\cal P}_s^{(\phi)} (k) + {\cal P}_s^{(\chi)} (k))}{d \ln k},
\end{equation}
with which we can describe the power spectrum as if there is only one power spectrum. 

To explicitly express the effective spectral index and its runnings with the slow-roll parameters, 
we also need to define the fraction of the contribution to the (total) power spectrum from  $\phi$ and $\chi$ fields as
\begin{equation}
Q_\phi \equiv \frac{ {\cal P}_s^{(\phi)} (k_0) }{ {\cal P}_s^{(\phi)} (k_0) + {\cal P}_s^{(\chi)} (k_0)},
\qquad
Q_\chi \equiv \frac{ {\cal P}_s^{(\chi)} (k_0) }{ {\cal P}_s^{(\phi)} (k_0) + {\cal P}_s^{(\chi)} (k_0)},\label{eq:Qchi}
\end{equation}
where these quantities are to be evaluated at the pivot scale. 
In some literature, the ratio $R$ between ${\cal P}_s^{(\chi)}$ and ${\cal P}_s^{(\phi)}$ is also used to characterize 
the contribution from the spectator
\begin{equation}
\label{eq:def_R}
R \equiv \frac{{\cal P}_s^{(\chi)}(k_0) }{{\cal P}_s^{(\phi)} (k_0)}  \left( = \frac{Q_\chi}{1 - Q_\chi} \right),
\end{equation}
which is again defined at the pivot scale.
With these variables, the spectral index and its runnings are given by 
\begin{eqnarray}
\label{eq:ns_eff}
n_s^{\rm (eff)}  -1 = 
&& 
Q_\phi  (n_s^{(\phi)} -1) +  Q_\chi  (n_s^{(\chi)} - 1),
\\
\label{eq:alpha_eff}
\alpha_s^{\rm (eff)} = 
&& 
Q_\phi  \alpha_s^{(\phi)} +  Q_\chi  \alpha_s^{(\chi)} + Q_\phi Q_\chi (\Delta n_s)^2,
 \\ 
\label{eq:beta_eff}
\beta_s^{\rm (eff)} =  && 
Q_\phi  \beta_s^{(\phi)} +  Q_\chi  \beta_s^{(\chi)}  + 3 Q_\phi Q_\chi \Delta n_s \Delta \alpha_s
- Q_\phi Q_\chi   (Q_\phi - Q_\chi) (\Delta n_s)^3,
 \\
\gamma_s^{\rm (eff)} = 
&& 
Q_\phi  \gamma_s^{(\phi)} +  Q_\chi  \gamma_s^{(\chi)} 
+ Q_\phi Q_\chi ( 4 \Delta n_s \Delta \beta_s + 3 (\Delta \alpha_s)^2)  \notag \\
&&
- 6 Q_\phi Q_\chi   (Q_\phi - Q_\chi)  ( \Delta n_s)^2 \Delta \alpha_s 
+ \{ Q_\phi Q_\chi   (Q_\phi - Q_\chi)^2 - 2 Q_\phi^2 Q_\chi^2 \}   ( \Delta n_s)^4, \notag \\
 \\
\delta_s^{\rm (eff)}  = 
&& 
Q_\phi  \delta_s^{(\phi)} +  Q_\chi  \delta_s^{(\chi)} 
+ Q_\phi Q_\chi ( 10  \Delta \alpha_s \Delta \beta_s + 5 \Delta n_s \Delta \gamma_s)   \notag \\
&&
- Q_\phi Q_\chi   (Q_\phi - Q_\chi) \{ 10 ( \Delta n_s)^2 \Delta \beta_s  + 15 \Delta n_s (\Delta \alpha_s)^2 \} \notag \\
&&
+ \{10 Q_\phi Q_\chi ( Q_\chi - Q_\phi)^2 - 20 Q_\phi^2 Q_\chi^2 \}  ( \Delta n_s)^3 \Delta \alpha_s \notag \\
&&
-\{Q_\phi Q_\chi ( Q_\chi - Q_\phi)^3 - 8 Q_\phi^2 Q_\chi^2 ( Q_\chi - Q_\phi) \}  ( \Delta n_s)^5,
\label{eq:ns_runnings_multi}
\end{eqnarray}
where 
\begin{equation}
 \Delta n_s = n_s^{(\phi)} - n_s^{(\chi)},
\quad
 \Delta \alpha_s = \alpha_s^{(\phi)} - \alpha_s^{(\chi)},
\quad
 \Delta \beta_s = \beta_s^{(\phi)} - \beta_s^{(\chi)},
 \quad
 \Delta \gamma_s = \gamma_s^{(\phi)} - \gamma_s^{(\chi)},
 \quad
 \Delta \delta_s = \delta_s^{(\phi)} - \delta_s^{(\chi)}.
\end{equation}

In the limits where $Q_\phi \rightarrow 0$ and $Q_\chi \rightarrow 0$, 
the expressions become the same as the pure spectator and inflaton cases, respectively.

The tensor-to-scalar ratio in multi-field models is given by 
\begin{equation}
\label{eq:r_multi}
r^{\rm (multi)} = 16 \epsilon Q_\chi,
\end{equation}
from which one can see that the tensor-to-scalar ratio is generally suppressed in multi-field models.

\section{Constraints on higher order spectral runnings}
\label{app:higher_running}

\begin{figure}
  \begin{center}
    \hspace{0mm}\scalebox{2.}{\includegraphics{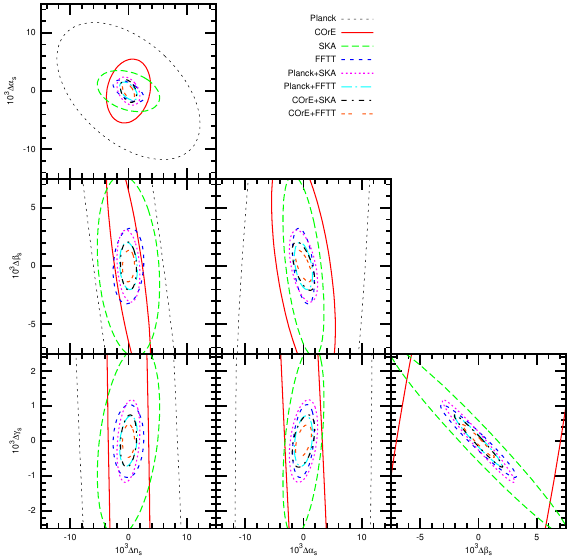}}
  \end{center}
    \vspace{10mm}
  \caption{
  Constraints on the spectral index $n_s$ and its runnings up to the cubic order.
  For 21~cm line observations, $z_{\rm min}=6$ is assumed.
  }
  \label{fig:zmin06_abc}
\end{figure}

In the main text, we have truncated the expansion at the quadratic order running $\beta_s$. 
However, minihalos can probe small scale fluctuations at $20~{\rm Mpc}^{-1} < k < 500~{\rm Mpc}^{-1}$. 
Therefore, we can also probe the higher order runnings such as the cubic and quartic runnings,  $\gamma_s$ and $\delta_s$. 
In Fig.~\ref{fig:zmin06_abc}, the expected constraints on $n_s$ and the runnings for the case with  $z_{\rm min}=6$ are given 
and their $1\sigma$ sensitivities are summarized in Table~\ref{tab:const_abc}.
Also, as discussed in the text, the constraints depend on the minimum redshift. 
In Table~\ref{tab:zdep_abc}, the dependence of $1\sigma$ errors is also summarized 
for the combinations of Planck+SKA and COrE+FFTT.

\begin{table}
  \begin{center}
    \begin{tabular}{lcccc}
      \hline\hline
      & $10^3\Delta n_s$ & $10^3\Delta \alpha_s$ & $10^3\Delta \beta_s$ & $10^3\Delta \gamma_s$ \\
      \hline
	Planck & 11.3 & 10.8 & 37.4 & 32.1 \\
	COrE & 3.5 & 5.0 & 9.4 & 14.2 \\
	SKA & 4.9 & 3.3 & 7.2 & 2.5 \\
	FFTT & 2.4 & 1.7 & 3.0 & 0.96 \\
	Planck+SKA & 1.8 & 2.2 & 2.8 & 1.0 \\
	Planck+FFTT & 1.4 & 1.4 & 1.8 & 0.66 \\
	COrE+SKA & 1.2 & 1.7 & 1.9 & 0.69 \\
	COrE+FFTT & 0.95 & 1.2 & 1.2 & 0.43 \\
      \hline\hline
    \end{tabular}
  \end{center}
  \caption{
  Expected 1$\sigma$ sensitivities for the spectral index $n_s$ and its runnings up to the cubic order.
  For 21~cm line observations, $z_{\rm min}=6$ is assumed.}
 \label{tab:const_abc}
\end{table}
\begin{table}
  \begin{center}
    \begin{tabular}{llcccc}
      \hline\hline
      & $z_{\rm min}$ & $10^{3}\Delta n_s$ & $10^{3}\Delta \alpha_s$ & $10^{3}\Delta \beta_s$ & $10^{3}\Delta \gamma_s$ \\
      \hline
	\multirow{4}{*}{Planck+SKA} 
	& 4 & 1.5 & 1.6 & 2.3 & 0.78 \\
	& 6 & 1.8 & 2.2 & 2.9 & 1.1 \\
	& 8 & 2.4 & 3.2 & 3.2 & 1.3 \\
	& 10 & 3.7 & 4.7 & 4.5 & 1.9 \\
	\hline
	\multirow{4}{*}{COrE+FFTT} 
	& 4 & 0.86 & 1.1 & 1.1 & 0.38 \\
	& 6 & 0.95 & 1.3 & 1.3 & 0.43 \\
	& 8 & 1.1 & 1.4 & 1.3 & 0.45 \\
	& 10 & 1.2 & 1.6 & 1.4 & 0.48 \\
      \hline\hline
    \end{tabular}
  \end{center}
  \caption{
  Dependence of the constraints on $z_{\rm min}$ for Planck+SKA and COrE+FFTT.
  }
 \label{tab:zdep_abc}
\end{table}

\providecommand{\href}[2]{#2}\begingroup\raggedright


\begin{thebibliography}{100}


\bibitem{Adam:2015rua} 
  R.~Adam {\it et al.} [Planck Collaboration],
  Astron.\ Astrophys.\  {\bf 594}, A1 (2016)
  doi:10.1051/0004-6361/201527101
  [arXiv:1502.01582 [astro-ph.CO]].
  
\bibitem{Ade:2015xua} 
  P.~A.~R.~Ade {\it et al.} [Planck Collaboration],
  Astron.\ Astrophys.\  {\bf 594}, A13 (2016)
  doi:10.1051/0004-6361/201525830
  [arXiv:1502.01589 [astro-ph.CO]].
  
\bibitem{Alam:2016hwk} 
  S.~Alam {\it et al.} [BOSS Collaboration],
  [arXiv:1607.03155 [astro-ph.CO]].

\bibitem{Betoule:2014frx} 
  M.~Betoule {\it et al.} [SDSS Collaboration],
  Astron.\ Astrophys.\  {\bf 568}, A22 (2014)
  doi:10.1051/0004-6361/201423413
  [arXiv:1401.4064 [astro-ph.CO]].

\bibitem{Ade:2015lrj}
  P.~A.~R.~Ade {\it et al.} [Planck Collaboration],
  Astron.\ Astrophys.\  {\bf 594} (2016) A20
  doi:10.1051/0004-6361/201525898
  [arXiv:1502.02114 [astro-ph.CO]].

\bibitem{Huang:2015gca} 
  Q.~G.~Huang, S.~Wang and W.~Zhao,
  JCAP {\bf 1510}, no. 10, 035 (2015)
  doi:10.1088/1475-7516/2015/10/035
  [arXiv:1509.02676 [astro-ph.CO]].

\bibitem{Errard:2015cxa} 
  J.~Errard, S.~M.~Feeney, H.~V.~Peiris and A.~H.~Jaffe,
  JCAP {\bf 1603}, no. 03, 052 (2016)
  doi:10.1088/1475-7516/2016/03/052
  [arXiv:1509.06770 [astro-ph.CO]].
 
\bibitem{Alonso:2016xft} 
  D.~Alonso, J.~Dunkley, B.~Thorne and S.~N{\ae}ss,
  Phys.\ Rev.\ D {\bf 95}, no. 4, 043504 (2017)
  doi:10.1103/PhysRevD.95.043504
  [arXiv:1608.00551 [astro-ph.CO]]. 
  
\bibitem{Abazajian:2016yjj} 
  K.~N.~Abazajian {\it et al.} [CMB-S4 Collaboration],
  arXiv:1610.02743 [astro-ph.CO].
  
\bibitem{Barron:2017kuo} 
  D.~Barron {\it et al.},
  arXiv:1702.07467 [astro-ph.IM].  
  
 \bibitem{Kohri:2013mxa} 
  K.~Kohri, Y.~Oyama, T.~Sekiguchi and T.~Takahashi,
  JCAP {\bf 1310}, 065 (2013)
  doi:10.1088/1475-7516/2013/10/065
  [arXiv:1303.1688 [astro-ph.CO]].
  
  \bibitem{Munoz:2016owz} 
  J.~B.~Mu{\~n}oz, E.~D.~Kovetz, A.~Raccanelli, M.~Kamionkowski and J.~Silk,
  arXiv:1611.05883 [astro-ph.CO].  

\bibitem{Dent:2012ne} 
  J.~B.~Dent, D.~A.~Easson and H.~Tashiro,
  Phys.\ Rev.\ D {\bf 86}, 023514 (2012)
  doi:10.1103/PhysRevD.86.023514
  [arXiv:1202.6066 [astro-ph.CO]].

\bibitem{Cabass:2016giw} 
  G.~Cabass, A.~Melchiorri and E.~Pajer,
  Phys.\ Rev.\ D {\bf 93}, no. 8, 083515 (2016)
  doi:10.1103/PhysRevD.93.083515
  [arXiv:1602.05578 [astro-ph.CO]].
   
\bibitem{Iliev:2002gj} 
  I.~T.~Iliev, P.~R.~Shapiro, A.~Ferrara and H.~Martel,
  Astrophys.\ J.\  {\bf 572}, 123 (2002)
  doi:10.1086/341869
  [astro-ph/0202410].

\bibitem{Iliev:2002ms} 
  I.~T.~Iliev, E.~Scannapieco, H.~Martel and P.~R.~Shapiro,
  Mon.\ Not.\ Roy.\ Astron.\ Soc.\  {\bf 341}, 81 (2003)
  doi:10.1046/j.1365-8711.2003.06410.x
  [astro-ph/0209216].

\bibitem{Shimabukuro:2014ava} 
  H.~Shimabukuro, K.~Ichiki, S.~Inoue and S.~Yokoyama,
  Phys.\ Rev.\ D {\bf 90}, no. 8, 083003 (2014)
  doi:10.1103/PhysRevD.90.083003
  [arXiv:1403.1605 [astro-ph.CO]].

\bibitem{Sekiguchi:2014wfa} 
  T.~Sekiguchi and H.~Tashiro,
  JCAP {\bf 1408}, 007 (2014)
  doi:10.1088/1475-7516/2014/08/007
  [arXiv:1401.5563 [astro-ph.CO]].

\bibitem{Takeuchi:2013hza} 
  Y.~Takeuchi and S.~Chongchitnan,
  Mon.\ Not.\ Roy.\ Astron.\ Soc.\  {\bf 439}, no. 1, 1125 (2014)
  doi:10.1093/mnras/stu059
  [arXiv:1311.2585 [astro-ph.CO]].
	
\bibitem{Sekiguchi:2013lma} 
  T.~Sekiguchi, H.~Tashiro, J.~Silk and N.~Sugiyama,
  JCAP {\bf 1403}, 001 (2014)
  doi:10.1088/1475-7516/2014/03/001
  [arXiv:1311.3294 [astro-ph.CO]].

\bibitem{Chongchitnan:2012we} 
  S.~Chongchitnan and J.~Silk,
  Mon.\ Not.\ Roy.\ Astron.\ Soc.\  {\bf 426}, L21 (2012)
  doi:10.1111/j.1745-3933.2012.01315.x
  [arXiv:1205.6799 [astro-ph.CO]].

\bibitem{Tashiro:2013xra}
  H.~Tashiro, T.~Sekiguchi and J.~Silk,
  JCAP {\bf 1401} (2014) 013
  doi:10.1088/1475-7516/2014/01/013
  [arXiv:1310.4176 [astro-ph.CO]].

\bibitem{ska} 
{\tt  https://www.skatelescope.org}
  	
\bibitem{Tegmark:2008au} 
  M.~Tegmark and M.~Zaldarriaga,
  Phys.\ Rev.\ D {\bf 79}, 083530 (2009)
  doi:10.1103/PhysRevD.79.083530
  [arXiv:0805.4414 [astro-ph]].
	
\bibitem{Shapiro:1998zp} 
  P.~R.~Shapiro and I.~T.~Iliev,
  Mon.\ Not.\ Roy.\ Astron.\ Soc.\  {\bf 307}, 203 (1999)
  doi:10.1046/j.1365-8711.1999.02609.x
  [astro-ph/9810164].

\bibitem{Mo:1995cs} 
  H.~J.~Mo and S.~D.~M.~White,
  Mon.\ Not.\ Roy.\ Astron.\ Soc.\  {\bf 282}, 347 (1996)
  doi:10.1093/mnras/282.2.347
  [astro-ph/9512127].

\bibitem{Madau:1996cs} 
  P.~Madau, A.~Meiksin and M.~J.~Rees,
  Astrophys.\ J.\  {\bf 475}, 429 (1997)
  doi:10.1086/303549
  [astro-ph/9608010].
	
\bibitem{Shapiro:2003gxa} 
  P.~R.~Shapiro, I.~T.~Iliev and A.~C.~Raga,
  Mon.\ Not.\ Roy.\ Astron.\ Soc.\  {\bf 348}, 753 (2004)
  doi:10.1111/j.1365-2966.2004.07364.x
  [astro-ph/0307266].
	
\bibitem{Adam:2016hgk} 
  R.~Adam {\it et al.} [Planck Collaboration],
  doi:10.1051/0004-6361/201628897
  arXiv:1605.03507 [astro-ph.CO].

\bibitem{Barkana:2000fd} 
  R.~Barkana and A.~Loeb,
  Phys.\ Rept.\  {\bf 349}, 125 (2001)
  doi:10.1016/S0370-1573(01)00019-9
  [astro-ph/0010468].
	
\bibitem{Kaiser:1987qv} 
  N.~Kaiser,
  Mon.\ Not.\ Roy.\ Astron.\ Soc.\  {\bf 227}, 1 (1987).

\bibitem{Knox:1995dq}
  L.~Knox,
  Phys.\ Rev.\  D {\bf 52}, 4307 (1995)
  [arXiv:astro-ph/9504054].
	
\bibitem{Furlanetto:2006jb} 
  S.~Furlanetto, S.~P.~Oh and F.~Briggs,
  Phys.\ Rept.\  {\bf 433}, 181 (2006)
  [astro-ph/0608032].
	
\bibitem{Tegmark:1996bz} 
  M.~Tegmark, A.~Taylor and A.~Heavens,
  Astrophys.\ J.\  {\bf 480}, 22 (1997)
  doi:10.1086/303939
  [astro-ph/9603021].

\bibitem{Planck:2006aa} 
  J.~Tauber {\it et al.} [Planck Collaboration],
  astro-ph/0604069.

\bibitem{core} 
{\tt http://www.core-mission.org}

\bibitem{Martin:2013tda} 
  J.~Martin, C.~Ringeval and V.~Vennin,
  Phys.\ Dark Univ.\  {\bf 5-6}, 75 (2014)
  doi:10.1016/j.dark.2014.01.003
  [arXiv:1303.3787 [astro-ph.CO]].
  
\bibitem{Starobinsky:1980te} 
  A.~A.~Starobinsky,
  Phys.\ Lett.\  {\bf 91B}, 99 (1980).
  doi:10.1016/0370-2693(80)90670-X

\bibitem{Nariai:1971sv} 
  H.~Nariai and K.~Tomita,
  Prog.\ Theor.\ Phys.\  {\bf 46}, 776 (1971).
  doi:10.1143/PTP.46.776

\bibitem{Tomita:2016tcj} 
  K.~Tomita,
  arXiv:1603.07621 [gr-qc].

\bibitem{Enqvist:2001zp}
K.~Enqvist and M.~S.~Sloth,
Nucl.\ Phys.\ B {\bf 626}, 395 (2002)
[arXiv:hep-ph/0109214].

\bibitem{Lyth:2001nq}
D.~H.~Lyth and D.~Wands,
Phys.\ Lett.\ B {\bf 524}, 5 (2002)
[arXiv:hep-ph/0110002].

\bibitem{Moroi:2001ct}
T.~Moroi and T.~Takahashi,
Phys.\ Lett.\ B {\bf 522}, 215 (2001)
[Erratum-ibid.\ B {\bf 539}, 303 (2002)]
[arXiv:hep-ph/0110096].

\bibitem{Dvali:2003em}
  G.~Dvali, A.~Gruzinov, M.~Zaldarriaga,
  Phys.\ Rev.\  {\bf D69}, 023505 (2004).
  [astro-ph/0303591].

\bibitem{Kofman:2003nx}
  L.~Kofman,
    [astro-ph/0303614].

\bibitem{Langlois:2004nn}
  D.~Langlois and F.~Vernizzi,
  Phys.\ Rev.\  D {\bf 70}, 063522 (2004)
  [arXiv:astro-ph/0403258];

\bibitem{Lazarides:2004we} 
  G.~Lazarides, R.~R.~de Austri and R.~Trotta,
  Phys.\ Rev.\ D {\bf 70}, 123527 (2004)
  [hep-ph/0409335].

\bibitem{Moroi:2005kz}
  T.~Moroi, T.~Takahashi and Y.~Toyoda,
  Phys.\ Rev.\  D {\bf 72}, 023502 (2005)
  [arXiv:hep-ph/0501007];

\bibitem{Moroi:2005np}
  T.~Moroi and T.~Takahashi,
  Phys.\ Rev.\  D {\bf 72}, 023505 (2005)
  [arXiv:astro-ph/0505339];

\bibitem{Ichikawa:2008iq}
  K.~Ichikawa, T.~Suyama, T.~Takahashi and M.~Yamaguchi,
  Phys.\ Rev.\  D {\bf 78}, 023513 (2008)
  [arXiv:0802.4138 [astro-ph]].

\bibitem{Ichikawa:2008ne}
  K.~Ichikawa, T.~Suyama, T.~Takahashi and M.~Yamaguchi,
  Phys.\ Rev.\ D {\bf 78}, 063545 (2008)
  [arXiv:0807.3988 [astro-ph]].
  
  \bibitem{Suyama:2010uj}
  T.~Suyama, T.~Takahashi, M.~Yamaguchi and S.~Yokoyama,
  JCAP {\bf 1012}, 030 (2010)
  [arXiv:1009.1979 [astro-ph.CO]].

\bibitem{Fonseca:2012cj} 
  J.~Fonseca and D.~Wands,
  JCAP {\bf 1206}, 028 (2012)
  [arXiv:1204.3443 [astro-ph.CO]].
  
\bibitem{Enqvist:2013paa}
  K.~Enqvist and T.~Takahashi,
  JCAP {\bf 1310} (2013) 034
  [arXiv:1306.5958 [astro-ph.CO]].

\bibitem{Vennin:2015vfa} 
  V.~Vennin, K.~Koyama and D.~Wands,
  JCAP {\bf 1511}, 008 (2015)
  [arXiv:1507.07575 [astro-ph.CO]].
    
\bibitem{Fujita:2014iaa} 
  T.~Fujita, M.~Kawasaki and S.~Yokoyama,
  JCAP {\bf 1409}, 015 (2014)
  doi:10.1088/1475-7516/2014/09/015
  [arXiv:1404.0951 [astro-ph.CO]].

\bibitem{Enqvist:2015njy} 
  K.~Enqvist, T.~Sekiguchi and T.~Takahashi,
  JCAP {\bf 1604}, no. 04, 057 (2016)
  [arXiv:1511.09304 [astro-ph.CO]].


\bibitem{Ade:2015ava} 
  P.~A.~R.~Ade {\it et al.} [Planck Collaboration],
  Astron.\ Astrophys.\  {\bf 594}, A17 (2016)
  doi:10.1051/0004-6361/201525836
  [arXiv:1502.01592 [astro-ph.CO]].
  
  \bibitem{Lorenz:2007ze} 
  L.~Lorenz, J.~Martin and C.~Ringeval,
  JCAP {\bf 0804}, 001 (2008)
  doi:10.1088/1475-7516/2008/04/001
  [arXiv:0709.3758 [hep-th]].

\bibitem{Freese:1990rb} 
  K.~Freese, J.~A.~Frieman and A.~V.~Olinto,
  Phys.\ Rev.\ Lett.\  {\bf 65}, 3233 (1990).
  doi:10.1103/PhysRevLett.65.3233
  
\bibitem{Adams:1992bn} 
  F.~C.~Adams, J.~R.~Bond, K.~Freese, J.~A.~Frieman and A.~V.~Olinto,
  Phys.\ Rev.\ D {\bf 47}, 426 (1993)
  doi:10.1103/PhysRevD.47.426
  [hep-ph/9207245].
	
\end{thebibliography}
\end{document}